\documentclass[%
 aps,
 amsmath,amssymb,
 reprint,%
]{revtex4-2}

\usepackage{graphicx}
\usepackage{dcolumn}
\usepackage{bm}

\usepackage[utf8]{inputenc}
\usepackage[T1]{fontenc}
\usepackage{mathptmx}
\usepackage{etoolbox}
\usepackage{xcolor}
\usepackage{colortbl}
\usepackage{hyperref}
\usepackage{url}
\usepackage{makecell}
\usepackage{placeins}

\makeatletter
\def\@email#1#2{%
 \endgroup
 \patchcmd{\titleblock@produce}
  {\frontmatter@RRAPformat}
  {\frontmatter@RRAPformat{\produce@RRAP{*#1\href{mailto:#2}{#2}}}\frontmatter@RRAPformat}
  {}{}
}%
\makeatother
\begin{document}

\preprint{}

\title{Machine-Learning-Compatible CALPHAD-type Optimization from Phase Equilibria by Auto-differentiation}

\author{Wenhao Zhang}
\affiliation{International Center for Young Scientists (ICYS), National Institute for Materials Science (NIMS), Sengen 1-2-1, Tsukuba, Ibaraki 305-0047, Japan}
\email{ZHANG.Wenhao@nims.go.jp}
\author{Jean-Claude Crivello}
\affiliation{%
CNRS-Saint-Gobain-NIMS, IRL 3629, Laboratory for Innovative Key Materials and Structures (LINK), 1-1 Namiki, Tsukuba, 305-0044, Ibaraki, Japan
}
\author{Yusuke Matsuoka}
\author{Toshiyuki Koyama}
\author{Taichi Abe}
\affiliation{%
Research Center for Structure Materials, National Institute for Materials Science (NIMS), Sengen 1-2-1, Tsukuba, Ibaraki 305-0047, Japan
}

\date{\today}

\begin{abstract}
To accurately determine phase boundaries and phase transitions, thermodynamic models that describe phase free energies often have to be optimized based on experimentally observed phase equilibria. While different approaches exist for thermodynamic optimization, they are often implemented in ways that are not compatible with machine learning workflows requiring differentiable calculation of the loss function. In this work, we derive a phase-equilibrium loss function based on thermodynamic potentials that can be efficiently evaluated and enables gradient-based optimization by auto-differentiation in the \texttt{PyTorch} package. By minimizing this loss function, general thermodynamic model parameters can be optimized with respect to experimental phase-equilibria data. Using thermodynamic models in the CALculation of PHAse Diagram (CALPHAD) framework, we illustrate successful and efficient optimization in different systems, including ternary systems with more than 100 parameters. As the loss function is defined independently of the details of the thermodynamic models, it can be used to optimize machine learning thermodynamic models in general. In particular, we demonstrate top-down optimization of an atomistic potential from target phase equilibria.
\end{abstract}

\maketitle

\section{Introduction}

Phase diagrams are of fundamental importance in materials development, design and optimization because they describe phase stability and phase equilibria as functions of temperature, pressure and composition. \cite{porterPhaseTransformationsMetals2009,chewPhaseDiagramsWhy2023} Since it is often impractical to map out complete phase diagrams using discrete experimental measurements alone, computational approaches are essential for interpolating, assessing and predicting phase diagrams. \cite{liuThermodynamicsItsPrediction2023a,lukasComputationalThermodynamicsCalphad2007}

If some experimental observations are available, CALculation of PHAse Diagram (CALPHAD) assessment can be used to efficiently interpolate between experimental data to determine phase boundaries and predict phase diagrams. \cite{lukasComputationalThermodynamicsCalphad2007} In the CALPHAD approach, phenomenological free energy models are parameterized as a function of temperature, pressure and site-fractions. The parameters are then determined by optimization with respect to experimental observed phase equilibria and available thermodynamic measurements such as formation energies or heat capacities. While it can allow accurate description of phase diagrams, the extrapolation to poorly characterized or previously unassessed materials systems is often limited because these optimized models are primarily descriptive.

Without experimental data, phase diagrams can be predicted from first principles. \cite{liuFirstPrinciplesCalculationsCALPHAD2009} In a typical approach, the free energy of a phase can be calculated from energies and forces obtained by density functional theory (DFT) calculations, or more recently, by machine learning interatomic potentials (MLIPs). \cite{riebesellFrameworkEvaluateMachine2025,poulAutomatedGenerationStructure2025,zhuAcceleratingCALPHADbasedPhase2025} Harmonic or quasi-harmonic phonon calculations account for vibrational entropy. \cite{togoFirstPrinciplesPhonon2015} Configurational entropy can be approximated in the Bragg-Williams approximation or, more accurately, based on the cluster expansion method or Monte-Carlo methods. \cite{walleSelfdrivenLatticemodelMonte2002} With the computed free energies, phase diagrams can be subsequently calculated using appropriate thermodynamic models.
However, practical use of phase diagrams depends crucially on the accurate description of phase boundaries and phase-transition temperatures, which demands small errors in energies well below the $\mathrm{kJ/mol}$ scale ($1\,\mathrm{kJ/mol}\approx 10.3\,\mathrm{meV/atom}$). On the other hand, both DFT and MLIP free energies can have errors in the range of $\mathrm{kJ/mol}$. \cite{zhangAccuracyUncertaintyQuantification2026} As a result, DFT- or MLIP-predicted phase diagrams can deviate quantitatively from experimental results and the prediction results may not be attractive for practical use such as guiding experimental efforts. A route to accurate phase diagrams is therefore often to use first-principles or MLIP predictions as initial thermodynamic descriptions and refine them using experimental measurements, as in CALPHAD assessments. 

As modern thermodynamic workflows increasingly involve flexibly parameterized models, including machine-learning (ML) thermodynamic models and atomistic potentials from which phase free energies can be derived, extending CALPHAD-type thermodynamic optimization to general complex thermodynamic models is important and is the main motivation of the current work.
This would allow thermodynamic optimization to integrate information across multiple scales, from experimental phase equilibria to first-principles and atomistic-model predictions, and support more accurate predictions for new phase diagrams.
To make this possible, the optimization objective must be formulated independently of the specific parameterization of the thermodynamic model and must be compatible with ML-workflows. While different types of experimental data can be used in the CALPHAD-type optimization, we focus on phase-equilibrium data because they are among the most important and widely used targets and directly constrain phase boundaries, invariant reactions and transition temperatures in the phase diagram.

Optimization from phase equilibrium data is not straightforward. Existing CALPHAD optimization approaches include numerical-gradient, black-box and analytic gradient methods \cite{kunselmanAnalyticallyDifferentiableMetrics2024,kunselmanAnalyticalGradientBasedOptimization2025} that cannot be extended easily beyond the CALPHAD framework. Auto-differentiation may be the most promising route to enable gradient-based optimization of any thermodynamic model regardless of complexity. In 2022, Guan demonstrated an auto-differentiation-based thermodynamic optimization with a purpose-built loss function for Cu--Rh binary. \cite{guanDifferentiableThermodynamicModeling2022} Most recently, differentiable calculation of liquid--liquid equilibrium (LLE) has been used to learn free-energy models of molecular systems. \cite{benhichamDifferentiableThermodynamicPhaseEquilibriaForMachine2026} 
Here, we derive a phase-equilibrium loss function from equilibrium conditions and implement it as a differentiable computation graph in \texttt{PyTorch}, enabling gradient-based optimization of thermodynamic models with different parameterizations.

As the loss function is of key importance in thermodynamic optimization, we briefly provide an overview of loss functions that are used in the CALPHAD approach. One form of the loss function is the sum of squared errors \cite{caoPANDATSoftwarePanEngine2009,tangUsingPARROTModule2016,reisUserfriendlyRobustCalphad2025}:
$$
\mathcal{L} = \frac{1}{2}\sum_i w_i [e_i - \phi_i(\mathbb{W})]^2
$$
where $i$ indexes experimental data, and $w_i$ are the associated weights related to the reliability of the data. $e_i$ are experimentally observed values, such as equilibrium composition, and $\phi_i$ is the corresponding value calculated from parameters $\mathbb{W}$. This form is used, for example, in the \texttt{PARROT} program. \cite{tangUsingPARROTModule2016} However, such a loss function can be difficult to use, especially when values cannot be calculated due to poor conditioning. \cite{reisUserfriendlyRobustCalphad2025} Alternatively, loss functions based on thermodynamic quantities have been used. For example, a phase-equilibrium condition has been used as a loss function in \texttt{PanOptimizer} \cite{caoPANDATSoftwarePanEngine2009}:
$$
\mathcal{L} = \frac{1}{2} \sum_i w_i \sum_A \left[\frac{\mu_A^{\alpha}(\mathbb{W})-\mu_A^{\beta}(\mathbb{W})}{RT_i}\right]^2
$$
where $\mu_A^{\alpha/\beta}$ is the chemical potential of component $A$ in phase $\alpha/\beta$ and $T_i$ is the temperature. Thermodynamic driving forces \cite{sundmanImplementationAlgorithmCalculate2015} have also been used to define loss functions. The recent implementation of gradient-based optimization by Kunselman et al.\ uses the following loss function for a stable phase $\alpha$ in phase equilibrium $i$ \cite{bocklundESPEIEfficientThermodynamic2019,kunselmanAnalyticalGradientBasedOptimization2025}:
$$
l_{i,\alpha} = \sum_A \bar{\mu}_A X_A^{\alpha} - G^{\alpha}
$$
where $\bar{\mu}_A$ is the chemical potential of $A$ in a defined target hyperplane and $G^{\alpha}$ is the Gibbs energy at the observed composition. The target hyperplane is obtained from global equilibrium calculations. Finally, the total loss is written as a sum of squares of the loss for each observed stable phase. While minimizing this loss to zero is a necessary and sufficient condition for reproducing observed phase equilibria, the computation of chemical potential is not straightforward using differentiable calculation. The loss function we derived below, on the other hand, can be calculated differentiably and efficiently.

\section{Results}

\subsection{Overview of the optimization}

\begin{figure}[ht!]
\centering
\includegraphics[width=\linewidth]{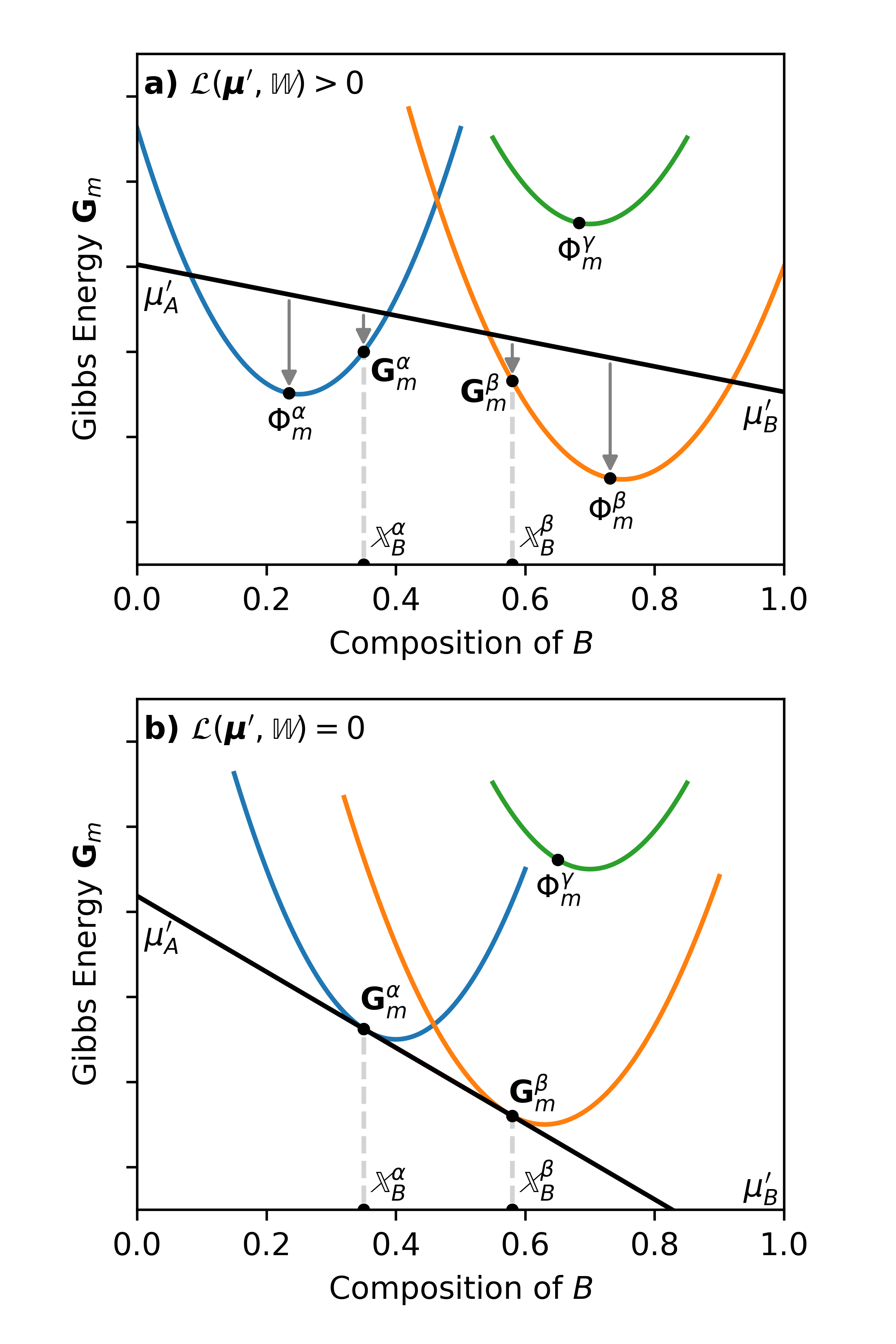}
\caption{Illustration of thermodynamic quantities used in the loss function in a binary system. 
Upper panel: before optimization when the loss $\mathcal{L}(\boldsymbol{\mu}',\mathbb{W})>0$. 
Lower panel: when the loss $\mathcal{L}(\boldsymbol{\mu}',\mathbb{W})=0$, where the experimentally observed phases and their compositions are reproduced by the optimized thermodynamic model.
$\mu_A'$ and $\mu_B'$ are the auxiliary chemical potentials. $\mathbb{X}_B^{\alpha/\beta}$ are the target atomic fraction of element $B$ for the phase $\alpha$ and $\beta$ in equilibrium and $\mathbf{G}_m$ are the Gibbs energies. $\gamma$ is an unobserved phase. $\phi_m$ denotes the energy difference between the phases and the chemical potential plane.}
\label{fig:loss}
\end{figure}

We first construct a differentiable loss function that measures whether a thermodynamic model reproduces a target phase equilibrium. Its derivation is detailed in the Methods section and follows from equilibrium conditions. The key idea is to test whether the experimentally observed phases can coexist under a common auxiliary chemical-potential hyperplane, while all unobserved phases remain unstable. This avoids repeated global equilibrium calculations during training and makes the objective compatible with auto-differentiation. The derived loss function is:
\begin{align*}
\mathcal{L}(\boldsymbol{\mu}',\mathbb{W}) &= \sum_{\alpha\in\text{stable}} \left|\mathbf{G}_m^{\alpha}(\mathbb{X}^{\alpha}, \mathbb{W}^{\alpha}) - \sum_A \mu'_A \mathbb{X}_{A}^{\alpha}\right|^p \\ &+ 
\sum_{\alpha\in\text{stable}} \left|\min_{\mathbf{y}^{\alpha},C_k^{\alpha}} \Phi_m^{\alpha}(\mathbf{y}^{\alpha},\boldsymbol{\mu}',\mathbb{W}^{\alpha})\right|^p \\ &+ \sum_{\gamma\notin\text{stable}} \mathrm{ReLU}\left[-\min_{\mathbf{y}^{\gamma},C_k^{\gamma}} \Phi_m^{\gamma}(\mathbf{y}^{\gamma},\boldsymbol{\mu}',\mathbb{W}^{\gamma})\right]
\end{align*}
where $\boldsymbol{\mu}'$ is the aforementioned auxiliary chemical potential, with $\mu_A'$ being the molar chemical potential of atoms of species $A$. $\mathbb{W}$ are the parameters of the thermodynamic models to be optimized. $p$ is the exponent of the loss and we use $p=1$ in this work, which gives faster convergence at small losses. A target phase equilibrium is reproduced when all three terms of the loss function are zero. A geometric interpretation of the terms in the loss function is shown in Figure~\ref{fig:loss}. The first term locally constrains the equilibrium solution to occur at the experimentally observed compositions, denoted by $\mathbb{X}_{A}^{\alpha}$ for species $A$ in phase $\alpha$. $\mathbf{G}_m^{\alpha}(\mathbb{X}^{\alpha}, \mathbb{W}^{\alpha})$ is the molar Gibbs energy of phase $\alpha$ at this constrained composition. In the second and third terms, we use a grand-potential-like quantity:
$$
\Phi_m^{\alpha}(\mathbf{y}^{\alpha}, \boldsymbol{\mu}', \mathbb{W}^{\alpha})
= G_m^{\alpha}(\mathbf{y}^{\alpha}, \mathbb{W}^{\alpha}) - \sum_A \mu_A' X_{A}^{\alpha}(\mathbf{y}^{\alpha})
$$
where $G_m^{\alpha}(\mathbf{y}^{\alpha}, \mathbb{W}^{\alpha})$ is the thermodynamic model describing the free energy of phase $\alpha$ as a function of a set of internal coordinates $\mathbf{y}^{\alpha}$. These internal coordinates may be required to satisfy a set of constraints $C_k^{\alpha}(\mathbf{y}^{\alpha}) = 0$. $X_{A}^{\alpha}(\mathbf{y}^{\alpha})$ is the atomic fraction of $A$ in phase $\alpha$ as a function of the same internal coordinates. The minimization $\min_{\mathbf{y}^{\alpha},C_k^{\alpha}}$ is with respect to $\mathbf{y}^{\alpha}$ subject to the constraints $C_k^{\alpha}$. On the other hand, the Gibbs energy $\mathbf{G}_m^{\alpha}(\mathbb{X}^{\alpha}, \mathbb{W}^{\alpha})$ can be obtained from $\min_{(\mathbf{y}^{\alpha}, C_k^{\alpha}, \mathbb{X}_{A}^{\alpha})} G_m^{\alpha}(\mathbf{y}^{\alpha}, \mathbb{W}^{\alpha})$. Following these definitions, the second term in the loss function ensures that the observed phases appear in the equilibrium solution. The third term ensures that unstable phases do not appear.
These two terms, however, do not specify where in composition space the observed phases appear. The first term provides this local constraint.

Optimal parameters $\mathbb{W}_{\mathrm{opt}}$ can be found as:
$$
\mathbb{W}_{\mathrm{opt}}= \arg\min_{\mathbb{W}} \left[ \sum_{t}\min_{\boldsymbol{\mu}_{t}'} \mathcal{L}_t(\boldsymbol{\mu}_t',\mathbb{W}) + \lambda \|\mathbb{W}\|_2^2 \right]
$$
where we have introduced regularization with weight $\lambda$. The first term includes all target equilibria indexed by $t$ and should be zero when all experimentally observed phase equilibria are reproduced. In practice, $\mathbb{W}_{\mathrm{opt}}$ can be found by minimizing $\mathcal{L}(\boldsymbol{\mu}',\mathbb{W})+ \lambda \|\mathbb{W}\|_2^2$ with respect to $\mathbb{W}$ and $\boldsymbol{\mu}'$ at the same time. 
The optimization with respect to the auxiliary chemical potential is analogous to phase-equilibrium calculations in chemical-potential space. \cite{fuClusterbasedComputationalThermodynamics2024}
We also note that the first term provide a relationship between Gibbs energy and the auxiliary chemical potential $\boldsymbol{\mu}'$, and can be used to eliminate some of the $\boldsymbol{\mu}'$s. The treatment of the auxiliary chemical potential and implementation of the calculation of $\mathcal{L}$ using the envelope theorem are discussed in the Methods section.

The inputs to the optimization are a set of defined thermodynamic models $G_m^{\alpha}(\mathbf{y}^{\alpha}, \mathbb{W}^{\alpha})$ with parameters to be optimized and the input phase-equilibria data. These phase equilibria can be separated into training, validation and test data. Training data are used to optimize thermodynamic parameters, validation data can be used to optimize hyperparameters, and test data can be used to assess the final optimization error. In the following examples, we use all input data for training for simplicity and judge the success of optimization based on final loss and the resulting phase diagrams. The Adam optimizer is used for gradient descent optimization. 
We have run optimization for $1000$ epochs and we have set learning rate for each system separately with a cosine decay to a minimum of $0.2$ of the initial value. 
Gradients are accumulated over all training data before being used to update the parameters. 

In the following, we use CALPHAD-type models as the main demonstration not only because they provide a controlled and established setting for thermodynamic optimization, but also because they play a central conceptual role by providing continuous free-energy descriptions of phases as functions of composition and temperature, which are necessary for phase-diagram determination. The loss function itself is independent of the model form, making it applicable to more general differentiable thermodynamic models. We therefore also demonstrate a proof-of-concept optimization of an embedded atom method potential from target phase equilibria.

\subsection{Optimization of CALPHAD-type thermodynamic models}

\begin{figure*}[!t]
\centering
\includegraphics[width=\linewidth]{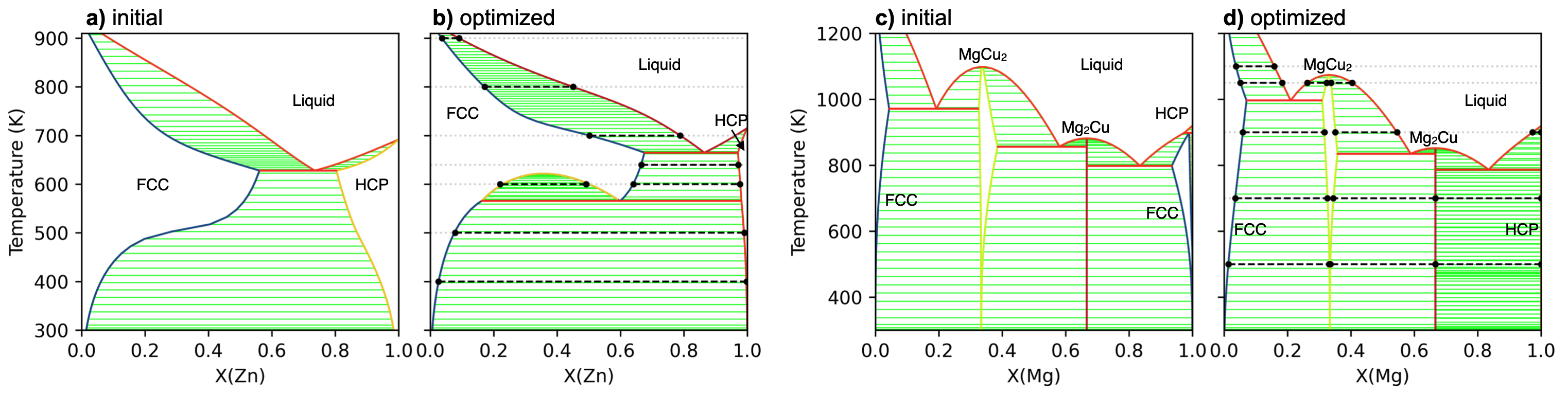}
\caption{Optimization results of Al--Zn and Cu--Mg binary systems. For the two binaries, the initial phase diagrams are shown on the left (a) and (c), and the optimized phase diagrams together with training phase-equilibria data are shown on the right (b) and (d). The target phase equilibria are shown as black dashed lines connecting two compositions.}
\label{fig:binary}
\end{figure*}

\begin{table*}
\caption{
Summary of All Optimization Examples, Including the Source of the Initial Model Before Optimization, the Number of Phases, the Total Number of Optimized Parameters, the Number of Training Phase Equilibria, Learning Rate (LR) and the Final Value of the Loss Function. The Loss Is Reported as $\mathcal{L}(\boldsymbol{\mu}',\mathbb{W})$ Divided by the Number of Training Phase Equilibria and Is in the Unit of $\mathrm{J/mol}$. 
}
\label{tab:table_summary}
\begin{ruledtabular}
\begin{tabular}{ccccccc}
System & Initial model & Phases & Parameters & Equilibria & LR & Final loss\\
\hline
Al--Zn & MLIP-derived  & 3 & 22 & 8 & 200 & 0.828 \\
Cu--Mg & MLIP-derived  & 9 & 38 & 13 & 200 & 0.451 \\
Al--Mo--Si & Perturbed from Ref.~\cite{liuThermodynamicReassessmentMo2000} & 14 & 110 &  119 & 1000 & 1.593 \\
Cu--Au & EAM (Ref.~\cite{golaEmbeddedAtomMethod2018}) & 4 & 42 & 25 & 0.005 & 0.254 \\
\end{tabular}
\end{ruledtabular}
\end{table*}

First, we apply the above differentiable loss function to the optimization of thermodynamic model used in the CALPHAD method. In CALPHAD, the free energy of a phase is commonly described using the compound energy formalism (CEF), in which crystallographic sites are represented by sublattices. \cite{lukasComputationalThermodynamicsCalphad2007} A detailed description of the CEF model is provided in Methods subsection on thermodynamic model. The free energy for a formula unit at a given site fraction $\mathbf{y}$ is given by:
$$
G(\mathbf{y},T) = \sum_{I} P_{I}(\mathbf{y}) g_I + RT \sum_s N_s \sum_{i=A}^{N} y_i^{(s)}\ln y_i^{(s)} + G^{\mathrm{ex}}(\mathbf{y},T)
$$
where the first term describes an energy surface by a weighted sum over different end-member configurations $I$. Each end-member is associated with a free energy $g_I$, and the weight $P_I$ is given by the corresponding site fractions. The second term is the configurational entropy in the site approximation, where $y_i^{(s)}$ is the occupancy of species $i$ on sublattice $s$ and $N_s$ is the multiplicity of sublattice $s$. $N$ is the total number of chemical species considered. Finally, $G^{\mathrm{ex}}(\mathbf{y},T)$ is the excess free energy term, which is further parameterized by interaction parameters $L$. Both the end-member energies $g$ and interaction parameters $L$ are parameterized as functions of temperature using polynomials $a+bT+\cdots$. The set of coefficients $a$, $b$ in the polynomials determines the thermodynamic model. The CEF model can be used to describe free energy in different phases including solid solutions (single sublattice), line compounds (single end-member) and intermetallic phases (multi-sublattice and multi-end-member phases).

In the following optimization, instead of updating all coefficients of $g$ and $L$ directly, we consider lower-order corrections to the $g$ and $L$ terms:
\begin{align*}
g_i(T) &= g_i^{\mathrm{init}}(T) + w_i + v_iT \\
L_j(T) &= L_j^{\mathrm{init}}(T) + w_j + v_j T
\end{align*}
where $g_i(T)$ and $L_j(T)$ are terms used in the CEF model, while $g_i^{\mathrm{init}}(T)$ and $L_j^{\mathrm{init}}(T)$ are the initially parameterized model values that are frozen as a reference surface. The loss function is minimized with respect to the correction parameters $\mathbf{w}$ and $\mathbf{v}$, initialized to zero. This form reduces the degrees of freedom in the optimization and is suitable when the initial thermodynamic models deviate not far from reproducing the target results.

\begin{figure*}[!t]
\centering
\includegraphics[width=0.9\linewidth]{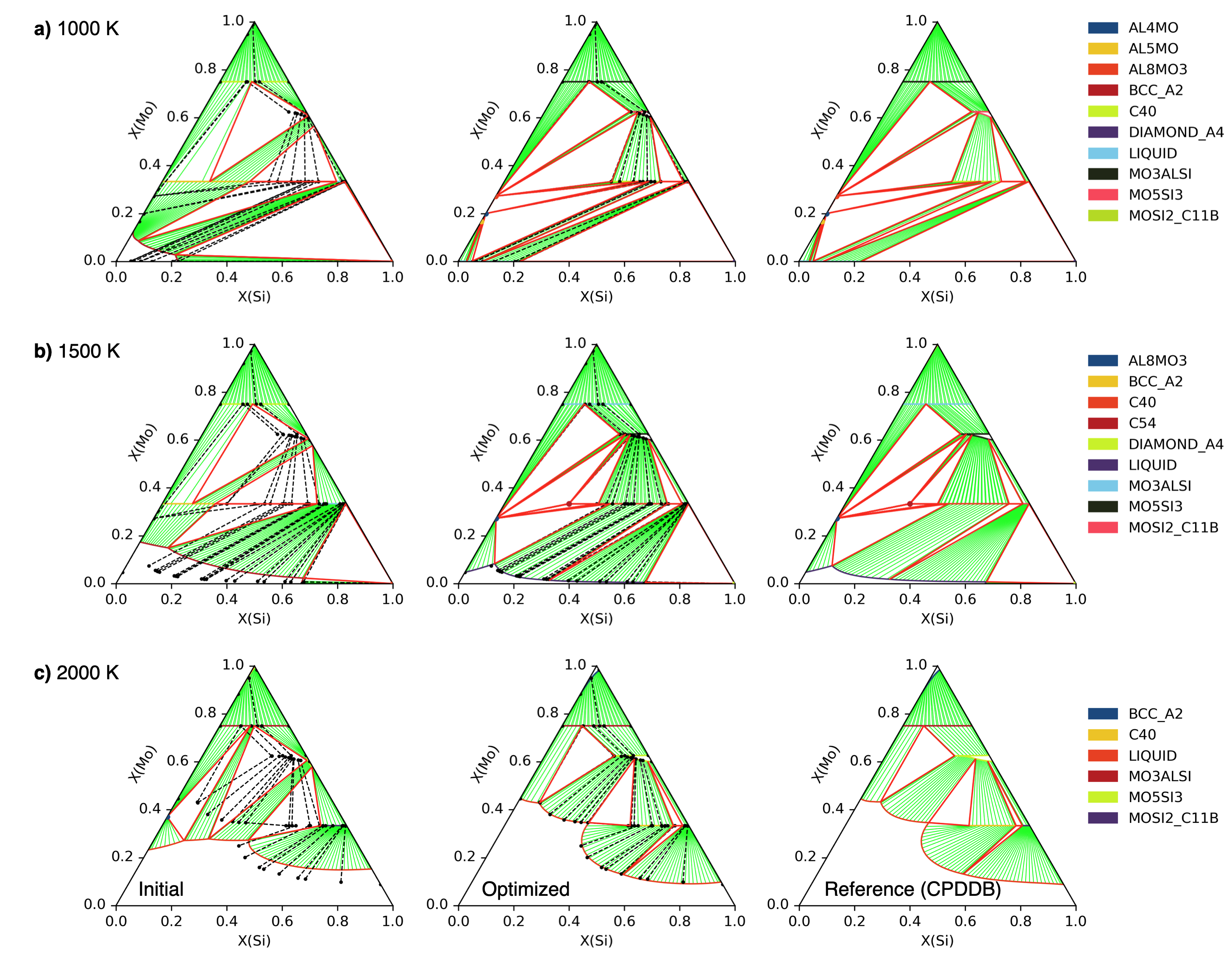}
\caption{Al--Mo--Si isothermal sections at three temperatures: (a) $1000\,\mathrm{K}$, (b) $1500\,\mathrm{K}$ and (c) $2000\,\mathrm{K}$. At each temperature, from left to right: phase diagrams from the initial thermodynamic models before optimization, phase diagrams after optimization, and the target phase diagram used to generate training data. Black dashed lines indicate phase equilibria used for training.}
\label{fig:ternary}
\end{figure*}

We apply the optimization to three example systems, Al--Zn, Cu--Mg and Al--Mo--Si, that contain solid solution phases, compounds and intermetallics. In these optimization examples, training phase-equilibria data are generated by equilibrium calculations from reference thermodynamic assessments hosted in the NIMS CPDDB database. \cite{abeCPDDB2007} For the Al--Zn \cite{meyReevaluationAlZnSystem1993} and Cu--Mg \cite{liangThermodynamicModellingCuMgZn1998a} binary systems, the initial thermodynamic models to be optimized are taken from our recent work and are obtained by fitting parameters to free energy calculations using universal MLIPs. \cite{zhangAccuracyUncertaintyQuantification2026} The NequIP and eSEN models are used for the Al--Zn and Cu--Mg binaries, respectively. \cite{batznerE3equivariantGraphNeural2022,fuLearningSmoothExpressive2025} For the ternary Al--Mo--Si system, we constructed a controlled optimization benchmark from the assessment of Y. Liu et al. \cite{liuThermodynamicReassessmentMo2000} The reference assessment was used as the target thermodynamic description, and the initial model was generated by perturbing parameters in the assessed models. This procedure gives an initial thermodynamic model with the same phase set and model topology as the target, but with phase equilibria that deviate substantially from the reference phase diagram. A summary is given in Table~\ref{tab:table_summary} and the thermodynamic model used for each phases are detailed in the \textbf{Supplementary information}. The initial phase diagrams before optimization are plotted using the \texttt{pycalphad} program in Figs.~\ref{fig:binary} and \ref{fig:ternary}. \cite{otisPycalphadCALPHADbasedComputational2017a}
Across these examples, the normalized final loss is reduced to below $2\,\mathrm{J/mol}$, indicating that the target phase-equilibria are closely satisfied after optimization (Table~\ref{tab:table_summary}). These final losses are small because the target phase equilibria are generated from consistent thermodynamic assessments. If experimental phase-equilibrium data are used, the final loss would be larger, reflecting the variance and uncertainty of the data.

\begin{figure}[t!]
\centering
\includegraphics[width=\linewidth]{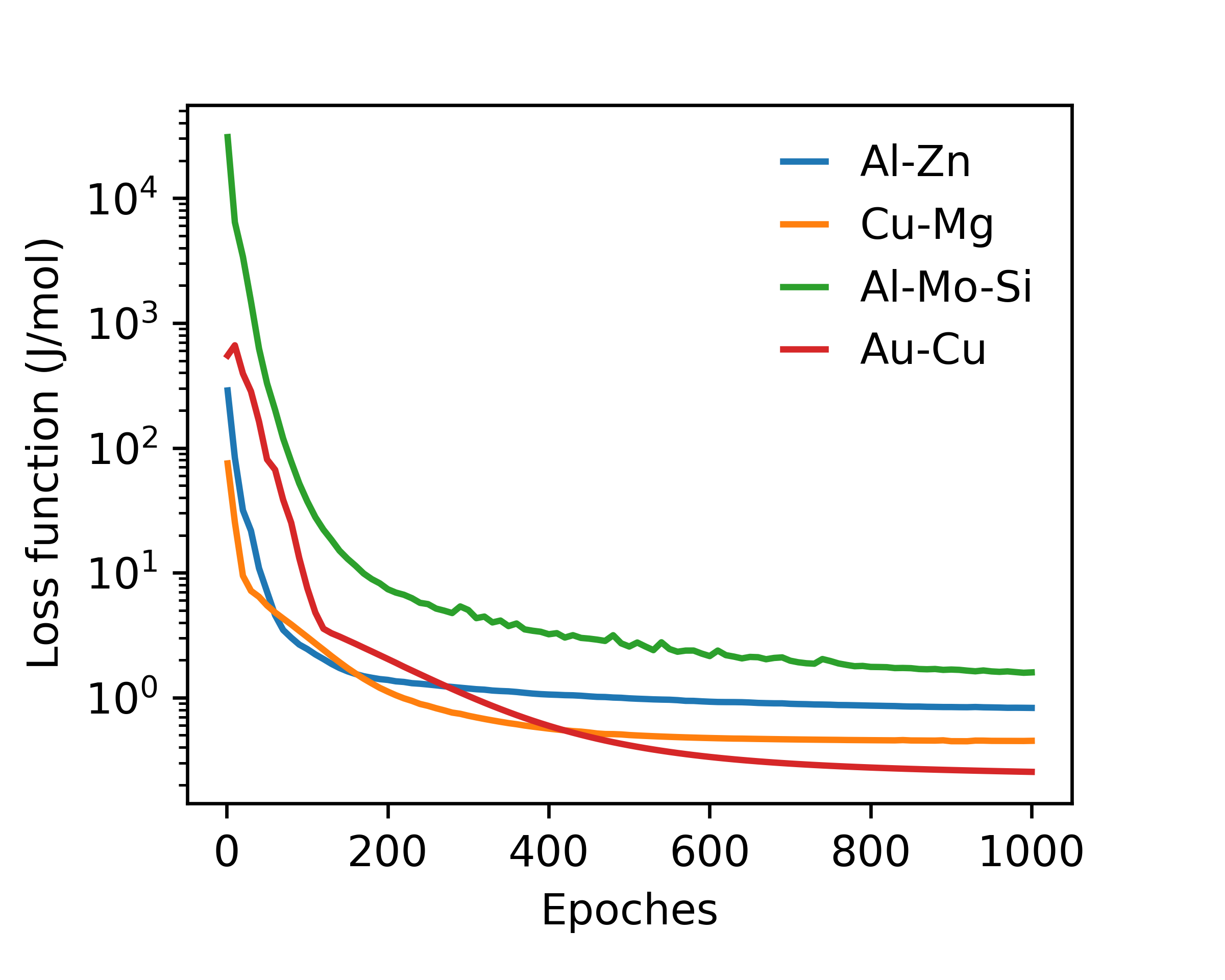}
\caption{Loss function, excluding the regularization term, plotted against the number of training epochs for Al--Zn, Cu--Mg, Al--Mo--Si and Cu--Au. The loss function is normalized by the number of phase equilibria used for training. $p=1$ is used to calculate the loss function; therefore, it has units of $\mathrm{J/mol}$.}
\label{fig:lc}
\end{figure}

The phase diagram of the Al--Zn binary consists of two solid solution phases and the liquid phase. At temperatures below $550\,\mathrm{K}$, a wide two-phase region between FCC and HCP can be found. At higher temperatures, the FCC phase extends from the Al-rich side up to $0.6$ atomic fraction of Zn, and a miscibility gap can be found in the assessed phase diagram. \cite{meyReevaluationAlZnSystem1993} The thermodynamic model before optimization (Fig.~\ref{fig:binary}a) correctly predicts a wide FCC region at high temperature but cannot reproduce the miscibility gap in the solid phase region. Furthermore, the homogeneity ranges of the HCP phase are severely overestimated. In the thermodynamic model, there are a total of $22$ parameters to be optimized. The training data consist of eight two-phase equilibria, as illustrated in Fig.~\ref{fig:binary}b, that are computed from $400\,\mathrm{K}$ to $900\,\mathrm{K}$ with a $100\,\mathrm{K}$ interval, with one additional phase equilibrium at $640\,\mathrm{K}$ just below the eutectic temperature. The corresponding learning curve is shown in Fig.~\ref{fig:lc}. The target phase equilibria are successfully reproduced after optimization corresponding to the minimized final loss (Fig.~\ref{fig:binary}b). 

The Cu--Mg system is a more difficult example because it involves two sublattices in the Laves phase MgCu$_2$ with internal degrees of freedom and a narrow composition range. Its phase diagram consists of five phases: the line compound Mg$_2$Cu, the Laves phase, the FCC phase on the Cu side, the HCP phase on the Mg side and the liquid phase. \cite{liangThermodynamicModellingCuMgZn1998a} In addition, four extra compound phases (MgCu$_3$, MgCu$_4$, MgCu, Mg$_5$Cu) taken from the Alexandria database are included in the optimization as unobserved phases. The Laves MgCu$_2$ phase is modelled using two sublattices from endmember energies. The initial phase diagram from MLIPs (Fig.~\ref{fig:binary}c) produces a narrow homogeneity region of the Laves phase, similar to assessment results, but shifted toward the Mg-rich side. Moreover, an additional FCC phase is stabilized on the Mg-rich side of the phase diagram due to MLIP error in the calculated energy. This initial thermodynamic description, with a total of $38$ parameters for correction, is used as the starting point of the optimization, and the optimized model reproduces the training phase-equilibria data well (Fig.~\ref{fig:binary}d). Notably, The optimization removes the spurious stabilization of the FCC phase and reproduces the homogeneity range of the Laves phase accurately.


The optimization example in the ternary Al--Mo--Si system is based on the assessment by Y. Liu et al. \cite{liuThermodynamicReassessmentMo2000} The initial phase diagrams at three different temperatures are plotted in the left-most column of Fig.~\ref{fig:ternary} and deviate substantially from the original assessment plotted in the right-most column of Fig.~\ref{fig:ternary}. It is a challenging example because the assessment contains 14 phases in the Al--Mo--Si system, four of which are solid solutions and four are modelled with multiple end-members. Again, we optimize corrections to the terms in CEF models, leading to 110 parameters to be optimized. In total, 119 training phase-equilibria data are generated at five different temperatures from $500\,\mathrm{K}$ to $2500\,\mathrm{K}$ with a step of $500\,\mathrm{K}$. The phase equilibria obtained at $1000\,\mathrm{K}$, $1500\,\mathrm{K}$ and $2000\,\mathrm{K}$ are shown in Fig.~\ref{fig:ternary}. The same approach as above is used to optimize the thermodynamic model with $1000$ gradient update steps, and the target phase equilibria across different temperatures are successfully reproduced. This example illustrates the accuracy and capability of the implementation in reproducing target phase equilibria even in complex multicomponent systems.

\begin{figure}[ht!]
\centering
\includegraphics[width=\linewidth]{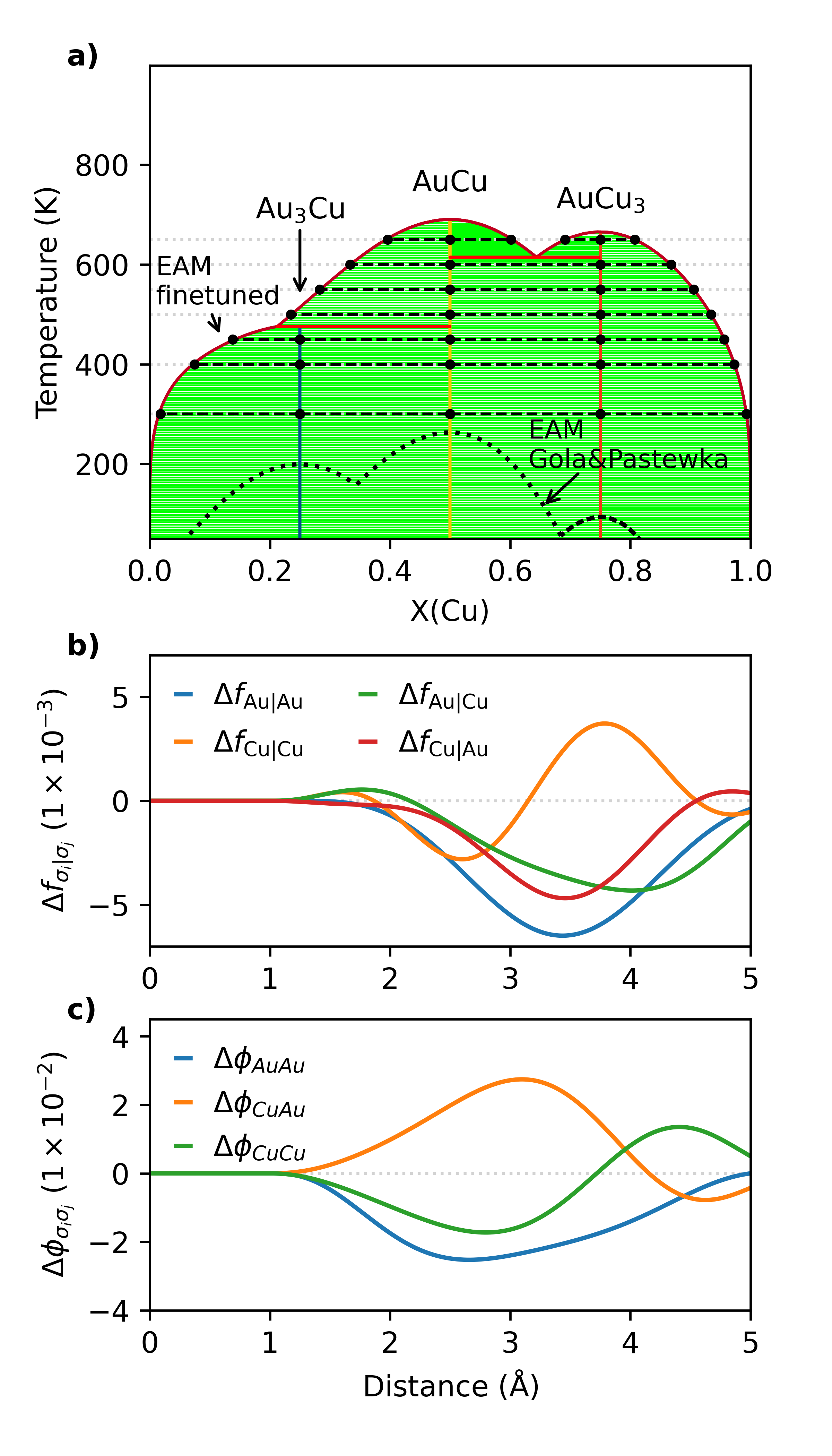}
\caption{Embedded atom method (EAM) potential optimized based on target phase equilibria in the Cu--Au binary. The target phase diagram is simplified by considering intermetallic phases as ordered compounds. The target phase equilibria are shown as black dashed lines connecting two compositions.
(a) Target phase equilibria used for training, and phase boundaries from the initial EAM model of Gola and Pastewka and the fine-tuned EAM model. 
(b) Optimized corrections to the density function in the EAM potential
(c) Optimized corrections to the pair-interaction function.}
\label{fig:eam}
\end{figure}

\subsection{EAM Potential in Au--Cu System}


To demonstrate that the proposed loss function can optimize parameters beyond CALPHAD-type descriptions, we consider a simplified Au--Cu binary model as a proof-of-concept example, in which an embedded atom method (EAM) potential is optimized from a set of target phase equilibria. 
The Au--Cu binary is a prototypical order-disorder transition system, and its phase diagram has been studied using both theoretical methods and thermodynamic assessments. Representative thermodynamic assessments include the work of Sundman using a four-sublattice split-CEF model \cite{sundmanThermodynamicAssessmentAuCu1998} and Cao et al.\ using the cluster site approximation. \cite{caoThermodynamicModelingCu2007} Theoretical phase diagrams of Au--Cu have also been computed from first-principles methods combined with CVM or Monte Carlo simulations. \cite{moritaClusterVariationMethod1957,weiFirstprinciplesCalculationsPhase1987,ozolinsCuAuAgAuCuAg1998}

Calculating the free energies of ordered and disordered phases in Au--Cu requires not only accurate energetics, including local relaxation effects, but also careful treatment of long-range and short-range ordering, with a relevant energy scale of $\mathrm{meV/atom}$. For our purpose, however, we simplify the thermodynamics of Au--Cu by introducing two important approximations:
first, the disordered phases are assumed to be ideal and the Bragg-Williams approximation is used for the configurational entropy. 
Second, we consider strictly ordered intermetallic phases with associated transition temperatures: $L1_0$ CuAu ($T=685\,\mathrm{K}$), $L1_2$ Cu$_3$Au ($T=665\,\mathrm{K}$) and its antistructure CuAu$_3$ ($T=475\,\mathrm{K}$). The three transition temperatures are derived from the review by Okamoto et al. \cite{okamotoAuCuGoldCopperSystem1987} As a consequence, this example does not aim to reproduce the experimental phase diagram. The target phase equilibria, derived from these transiton temperatures (see \textbf{Supplementary information}) are shown in Fig.~\ref{fig:eam} a. Vibrational entropies are ignored, and the Gibbs energies of the three intermetallic phases are described only by their formation enthalpies $H_{\mathrm{AuCu}}$, $H_{\mathrm{AuCu_3}}$ and $H_{\mathrm{Au_3Cu}}$ from EAM calculation. For the solid solution phase, the following expression is used for the Gibbs energy: 
\begin{align*}
G_{\mathrm{FCC}}(\mathbf{x},T) = & \sum_{i=\mathrm{Cu,Au}} x_i H_i + RT\sum_{i=\mathrm{Cu,Au}}x_i \ln x_i \\ &+ x_{\mathrm{Au}}x_{\mathrm{Cu}} \left( \sum_{n=0}^2 L_{\mathrm{Au,Cu}}^{(n)}(x_{\mathrm{Au}}-x_{\mathrm{Cu}})^n \right)
\end{align*}
where $H_{\mathrm{Cu/Au}}$ are also obtained from EAM calculations for FCC Cu and Au. The interaction parameters $L_{\mathrm{Au,Cu}}^{(n)}$ are temperature independent and are obtained by aligning the enthalpy part of $G_{\mathrm{FCC}}$ to the enthalpies of three 16-atom special quasirandom structures (SQS) at $x_{\mathrm{Cu}}$ equal to $0.25$, $0.5$ and $0.75$ that represent random solid solutions. Under the above simplification, $G_{\mathrm{FCC}}(\mathbf{x},T)$ is entirely parameterized from the enthalpies of five FCC structures.

The enthalpy values themselves are calculated from the EAM potential after full structural relaxation. The EAM potential is an efficient many-body potential in which the total energy is given by: \cite{golaEmbeddedAtomMethod2018}
$$
E = \sum_i \left[ F_{\sigma_i}\left(\sum_{j\neq i} f_{\sigma_i|\sigma_j}(r_{ij}) \right) + \frac{1}{2} \sum_{j\neq i} \phi_{\sigma_i\sigma_j}(r_{ij}) \right]
$$
where $i$ indexes a site and $j$ its neighboring sites. $\sigma_{i/j}$ is the chemical species (Cu and Au) on site $i/j$. $F_{\sigma_i}$ is the embedding function given for each chemical species. $f_{\sigma_i|\sigma_j}(r)$ is the density contribution from $\sigma_j$ at site $j$ to $\sigma_i$ at site $i$ when they are separated by distance $r$. Finally, $\phi_{\sigma_i\sigma_j}(r)$ is the pairwise interaction between $\sigma_i$ and $\sigma_j$, which is symmetric so that $\phi_{\sigma_i\sigma_j} = \phi_{\sigma_j\sigma_i}$. For the Cu--Au system, we used the potential developed by Gola and Pastewka, denoted $\text{EAM}_{\mathrm{GP}}$, which successfully reproduces many experimental quantities of the binary phases such as lattice parameters, elastic constants and bulk modulus. \cite{golaEmbeddedAtomMethod2018} However, as shown in Fig.~\ref{fig:eam} a and the phase diagram calculation in the original work, $\text{EAM}_{\mathrm{GP}}$ cannot accurately produce the transition temperature from ordered intermetallic phases to the solid solution phase.

We fine-tune the $\text{EAM}_{\mathrm{GP}}$ potential by optimizing additional correction terms $\Delta f(r)$ and $\Delta \phi(r)$ to the density function and the pairwise interaction so that the target phase-equilibrium loss is minimized. The distance-dependent correction is modelled by Chebyshev polynomials of the first kind, \cite{burdenNumericalAnalysis2011} which are then multiplied by an envelope function in an interval from $1\,\text{\r{A}}$ to $5\,\text{\r{A}}$, which is the cutoff distance of the EAM potential. The correction has the form:
$$
\Delta f(\mathbf{w},r)/\Delta \phi(\mathbf{w},r) = \left[ \sum_{i=0}^n w_n T_n(r) \right] \sin\left(\frac{r-r_{\mathrm{min}}}{r_{\mathrm{max}}-r_{\mathrm{min}}}\pi \right)
$$
where $T_n$ are Chebyshev polynomials of order $n$, $w_n$ are trainable coefficients and $n=6$ is found to be sufficient for the correction. The EAM potential with trainable corrections is implemented in \texttt{PyTorch} as an \texttt{ASE} calculator object with forces and stresses from auto-differentiation. The loss function is calculated as follows: i) for the EAM potential parameterized by $\mathbf{w}$, all eight structures are fully relaxed and the minimized energies as functions of trainable parameters $\mathbf{w}$ are obtained using the envelope theorem; ii) the parameters of the solid solution phase are obtained as functions of the minimized energies; and iii) phase-equilibrium loss is calculated from the Gibbs energies and grand potentials of all considered phases, as in the previous examples. The results of the optimization in terms of $\Delta f(r)$ and $\Delta \phi(r)$ are shown in Fig.~\ref{fig:eam} b and c. The optimized EAM potential satisfactorily reproduces the desired phase equilibrium (Fig.~\ref{fig:eam} a) and the learning curve is plotted in Fig.~\ref{fig:lc}.

Unlike typical CALPHAD optimization, in which different phases are parameterized by different parameters, all phases in this example share the same parameterization in terms of the underlying atomistic potential, which is optimized. Such atomistic potential fine-tuned from equilibrium data can be used not only to describe the phase diagram, as illustrated by the example, but also for other downstream tasks such as Monte Carlo or molecular dynamics simulations of the system. This is made possible by the flexible implementation of phase-equilibria optimization in a general ML-compatible environment.

\section{Discussion and conclusion}

Top-down optimization based on experimental data is an important approach in thermodynamic modeling of materials. For phase diagrams, the quantities being optimized are the free-energy surfaces of phases, which determine the phase equilibria that can be experimentally measured.
In this paper, we developed and demonstrated a ML-compatible optimization approach based on differentiable calculation of phase-equilibrium loss. Specifically, we have derived an ML-compatible loss function from phase-equilibrium condition and implemented it together with CALPHAD-style thermodynamic models. Through this framework, parameters that can determine thermodynamic free energies of phases can be optimized using gradient-based method. As the optimization uses auto-differentiation, it is not model-specific and can be used for parameters of the thermodynamic models as well as the parameters of underlying atomistic potentials. 

Using this new method, we demonstrated successful optimization of CALPHAD thermodynamic models in two binary systems Al--Zn and Cu--Mg and a challenging ternary system Al--Mo--Si toward target phase equilibria. In these optimizations, gradient descent drives the phase-equilibrium loss to within $2~\mathrm{J/mol}$ per phase equilibrium, and successfully reproduces the optimization targets. Furthermore, we have demonstrated the optimization of a simplified Au--Cu binary system in which the optimized parameters belong to an atomistic EAM potential. In this optimization, an end-to-end calculation from atomistic potential to phase-equilibrium loss is used through an intermediate thermodynamic model. While this only serves as a proof-of-concept, our method provides an ML-compatible link between thermodynamic models and phase equilibria, which can flexibly be used to optimize complex thermodynamic models as well as MLIPs. 

The top-down optimization of MLIPs based on experimental data is an important direction for potential development and our method based on phase equilibria offers another training target for top-down optimization. Recent MLIPs are mostly trained based on DFT data, with DFT error a fundamental bottleneck to MLIP accuracy in simulating and reproducing experimental results. Top-down optimization based on known experimental data therefore offers a route to partially correct systematic errors inherited from DFT and to align model predictions with experimental observables. Recent works have demonstrated the optimization of atomistic potentials based on pair distribution functions \cite{wangLearningPairPotentials2023}, elastic constants \cite{rockenAccurateMachineLearning2024} and phase transition temperatures \cite{fuchsRefiningMachineLearning2026}. Future work can combine phase-equilibrium losses with thermochemical and structural experimental targets, enabling multi-objective optimization of thermodynamic and atomistic models.

An important limitation of the current work is the omission of uncertainty quantification. Uncertainty-aware machine learning is important for top-down optimization for two reasons. First, the experimental targets themselves are uncertain because experimental data are often limited and contain measurement uncertainties. Second, simulation models, including thermodynamic models, contain approximations. Recently, Markov chain Monte Carlo (MCMC) and Hamiltonian Monte Carlo sampling of Bayesian posteriors have been implemented for CALPHAD optimization to enable uncertainty-aware modeling. \cite{paulsonQuantifiedUncertaintyThermodynamic2019,kunselmanGradientInformedHamiltonianMonte2026} Incorporating uncertainty into ML-based thermodynamic top-down optimization, including the approach developed here, is an important direction for future work.

\section{Methodology}

\subsection*{Equilibrium Conditions}
\label{eq_condition}

In this work, we use Greek letters $\alpha$, $\beta$ to indicate phases. A thermodynamic model $G_M^{\alpha}(\mathbf{y}^{\alpha}, T, P,\mathbb{W}^{\alpha})$ for a phase $\alpha$ is a function with which energy can be calculated from input variables $T$, $P$ and a set of internal coordinates $\mathbf{y}^{\alpha}$, which may be required to satisfy a set of constraints $C_k^{\alpha}(\mathbf{y}^{\alpha}) = 0$. Such internal coordinates can be sublattice site fractions in the CALPHAD framework or other order parameters depending on the thermodynamic model. The parameters of the model are denoted as $\mathbb{W}^{\alpha}$.
The subscript $M$ in $G_M$ indicates that $G_M$ is defined for one mole of the cell, in which vacancies may occupy some sites. The amount of chemical species $(A,B,\cdots)$ in one mole of the same cell is given as a function of internal coordinates:
$$
N_{A}^{\alpha} = N_{A}^{\alpha}(\mathbf{y}^{\alpha});\quad
N_{B}^{\alpha} = N_{B}^{\alpha}(\mathbf{y}^{\alpha});\quad\cdots
$$
The total amount of chemical species is $N^{\alpha} = \sum_{i\neq \mathrm{Vac}} N_i^{\alpha}$, and from this, we can calculate thermodynamic properties per atom: $G_m^{\alpha} = G_M^{\alpha}/ N^{\alpha}$.

We use black-bold letters to denote independent control variables. We consider a set of phases $\alpha, \beta, \gamma, \delta\cdots$ with phase fractions $\mathcal{N}^{\alpha}, \mathcal{N}^{\beta}\cdots$.
The global equilibrium at a given temperature $\mathbb{T}$, pressure $\mathbb{P}$ and total number of atoms $\mathbb{N}_A$ for each element $A$ can be obtained by minimizing the thermodynamic model subject to the constraints: \cite{sundmanImplementationAlgorithmCalculate2015}
\begin{align*}
\mathbf{G}(\mathbb{N}, \mathbb{T}, \mathbb{P}) &= \min_{(\mathcal{N}\ge 0,\mathbf{y},T,P)} \left[\sum_{\alpha} \mathcal{N}^{\alpha} G_M^{\alpha}(\mathbf{y}^{\alpha}, T, P,\mathbb{W}^{\alpha})\right] \\ &\text{subject to}
\begin{cases}
T = \mathbb{T}\\
P = \mathbb{P}\\
\sum_{\alpha}\mathcal{N}^{\alpha} N_{A}^{\alpha}(\mathbf{y}^{\alpha}) = \mathbb{N}_A, \cdots\\
C_1^{\alpha}(\mathbf{y}^{\alpha}) = 0, \cdots
\end{cases}
\end{align*}
It is important to note that we also have the inequality constraint $\mathcal{N}\ge 0$, so that for a phase $\gamma$ that is not represented in equilibrium, $\mathcal{N}^{\gamma}=0$. The minimization problem can be solved by finding the stationary point of the Lagrangian:
\begin{widetext}
\begin{align*}
L &=\sum_{\alpha} \mathcal{N}^{\alpha} G_M^{\alpha}(\mathbf{y}^{\alpha}, \mathbb{T}, \mathbb{P},\mathbb{W}^{\alpha})
+ \sum_A \mu_A \left[ \mathbb{N}_A-\sum_{\alpha}\mathcal{N}^{\alpha} N_{A}^{\alpha}(\mathbf{y}^{\alpha}) \right] + \sum_{\alpha} \sum_k \zeta_k^{\alpha} C_k^{\alpha}(\mathbf{y}^{\alpha}) \\
& = \sum_{\alpha} \mathcal{N}^{\alpha} \Phi_M^{\alpha}(\mathbf{y}^{\alpha}, \boldsymbol{\mu}, \mathbb{T}, \mathbb{P},\mathbb{W}^{\alpha})
+ \sum_A \mu_A \mathbb{N}_A + \sum_{\alpha} \sum_k \zeta_k^{\alpha} C_k^{\alpha}(\mathbf{y}^{\alpha})
\end{align*}
\end{widetext}
where $T=\mathbb{T}$ and $P=\mathbb{P}$ are solved trivially and can be inserted into the Lagrangian. They are dropped in the following. $\boldsymbol{\mu}$ and $\boldsymbol{\zeta}$ are Lagrange multipliers for the mass-balance constraints and constraints on $\mathbf{y}$. $\boldsymbol{\mu}$ can be shown to be the chemical potential. \cite{hillertViewpointsUseComputer1981} We have defined a quantity $\Phi$, which has the form of a grand potential because it is a function of the chemical potential:
$$
\Phi_M^{\alpha}(\mathbf{y}^{\alpha}, \boldsymbol{\mu}, \mathbb{W}^{\alpha})
= G_M^{\alpha}(\mathbf{y}^{\alpha}, \mathbb{W}^{\alpha}) - \sum_A \mu_A N_{A}^{\alpha}(\mathbf{y}^{\alpha})
$$
The stationary point is given by the following set of equations:
\begin{gather*}
\frac{\partial L}{\partial \mathcal{N}^{\alpha}} = \Phi_M^{\alpha}(\mathbf{y}^{\alpha}, \boldsymbol{\mu}, \mathbb{W}^{\alpha}) \begin{cases}=0\quad \text{if $\mathcal{N}^{\alpha}>0$}\\
> 0 \quad \text{if $\mathcal{N}^{\alpha}=0$} \end{cases} \\
\frac{\partial L}{\partial y_i^{\alpha}} = \mathcal{N}^{\alpha} \frac{\partial \Phi_M^{\alpha}}{\partial y_i^{\alpha}} + \sum_k \zeta_k \frac{\partial C_k^{\alpha}(\mathbf{y}^{\alpha})}{\partial y_i} = 0 \\
\frac{\partial L}{\partial \mu_A}=\sum_{\alpha}\mathcal{N}^{\alpha} N_{A}^{\alpha}(\mathbf{y}^{\alpha}) - \mathbb{N}_A = 0 \\
\frac{\partial L}{\partial \zeta_k^{\alpha}}=C_k^{\alpha}(\mathbf{y}^{\alpha}) = 0
\end{gather*}

where the above equations are defined for each phase, each internal coordinate, each chemical species and each constraint, respectively.
We note that for a stable phase with $\mathcal{N}^{\alpha}>0$, we require $\Phi_M^{\alpha}(\mathbf{y}^{\alpha}, \boldsymbol{\mu}, \mathbb{W}^{\alpha})=0$, but for an unstable phase $\gamma$, this condition does not need to be satisfied and the internal coordinates for phases with $\mathcal{N}=0$ cannot be uniquely determined. Nonetheless, with the determined chemical potential $\boldsymbol{\mu}$, the condition for a phase $\gamma$ to be unstable is that:
$$
\min_{\mathbf{y}^{\gamma},C_k^{\gamma}} \Phi_M^{\gamma}(\mathbf{y}^{\gamma},\mu,\mathbb{W}^{\gamma}) > 0
$$
where $C_k^{\gamma}$ means that the constraints on $\mathbf{y}^{\gamma}$ need to be satisfied.
On the other hand, for a stable phase $\alpha$, if we determine the same quantity $\min_{\mathbf{y}^{\alpha},C_k^{\alpha}} \Phi_M^{\alpha}(\mathbf{y}^{\alpha},\mu,\mathbb{W}^{\alpha})$, solving the minimization leads to the following equation, which is identical to the equilibrium condition:
$$
\frac{\partial \Phi_M^{\alpha}}{\partial y_i^{\alpha}} + \sum_k \zeta_k \frac{\partial C_k^{\alpha} (\mathbf{y}^{\alpha})}{\partial y_i^{\alpha}} = 0
$$
Thus, the internal coordinates $\mathbf{y}^{\alpha}$ solved from the global phase equilibrium also minimize $\Phi_M^{\alpha}$. Therefore, we can combine the above equilibrium conditions to give the equilibrium condition for a stable phase:
$$
\min_{\mathbf{y}^{\alpha},C_k^{\alpha}} \Phi_M^{\alpha}(\mathbf{y}^{\alpha},\mu,\mathbb{W}^{\alpha})=0
$$
A closely related grand-potential formulation is commonly used for phase-equilibrium calculations in chemical-potential space, where coexistence is identified from equal minimized grand potentials of competing phases at a common chemical potential. \cite{kikuchiTernaryPhaseDiagram1977,fuClusterbasedComputationalThermodynamics2024}

\subsection*{Loss Function}
\label{lossfunc}

\begin{figure}[ht!]
\centering
\includegraphics[width=\linewidth]{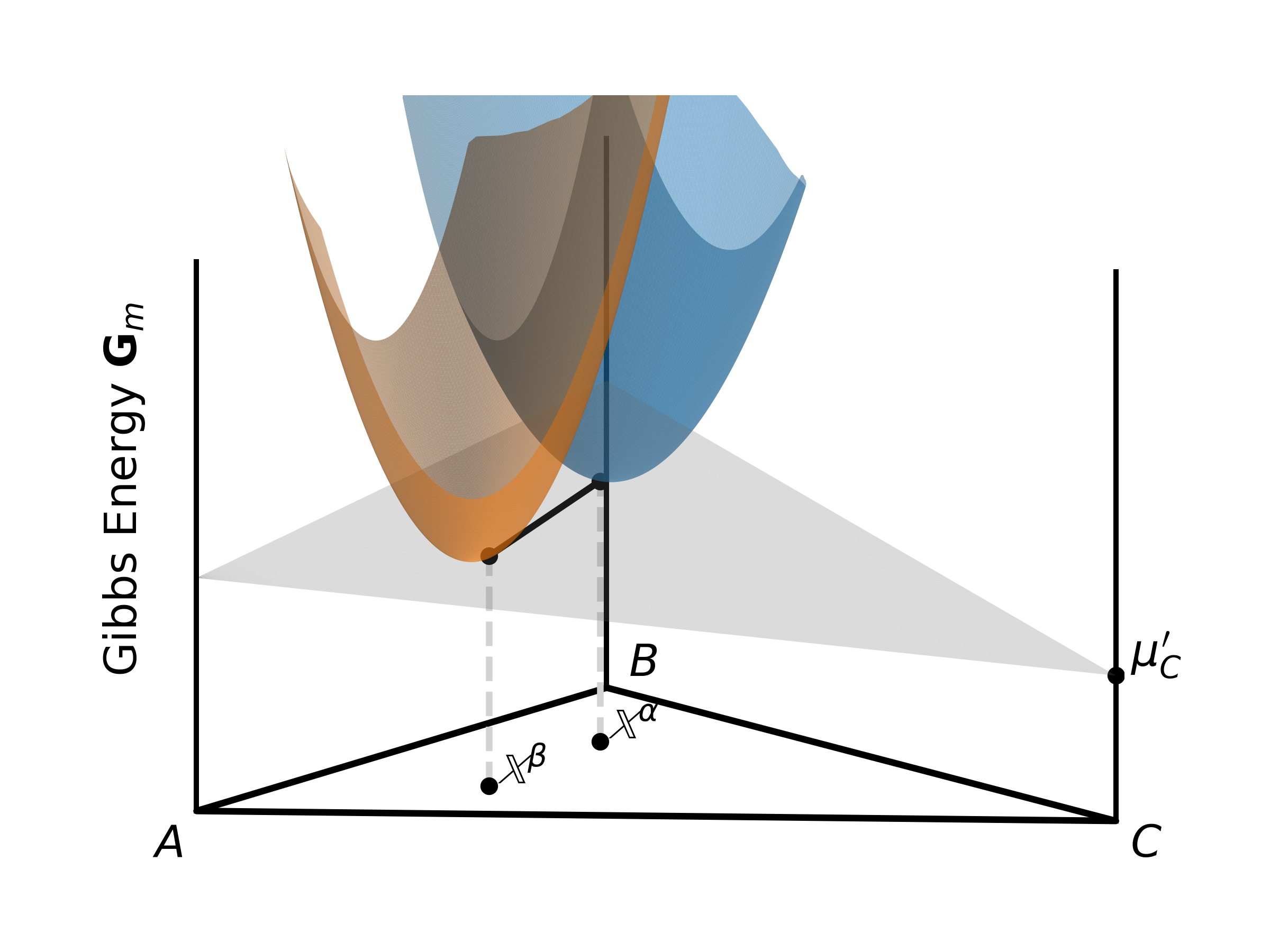}
\caption{Auxiliary chemical potential from a two-phase equilibrium in a ternary system. The auxiliary chemical potential requires one additional parameter.}
\label{fig:auxiliary}
\end{figure}

Typically, phase equilibria data are given by the measured compositions of phases in equilibrium: 
$$
(\mathbb{N}_A^{\alpha},\mathbb{N}_B^{\alpha},\cdots),
(\mathbb{N}_A^{\beta},\mathbb{N}_B^{\beta},\cdots),
\cdots
$$ 
If an equilibrium is indeed reproduced by the thermodynamic model, then the internal coordinates that satisfy the equilibrium conditions should also satisfy:
$$
N_A^{\alpha}(\mathbf{y}^{\alpha}) = \mathbb{N}_A^{\alpha}, \cdots
$$
Conversely, deviation from equilibrium can also be measured using the equilibrium condition, but at the constrained composition of each observed phase. Since the chemical potential is also unknown, we set auxiliary chemical potential vectors $\boldsymbol{\mu}'$ for this phase equilibrium and find the deviation from the observed phase equilibrium:
$$
\min_{\mathbf{y}^{\alpha},C_k^{\alpha}} \Phi_M^{\alpha}(\mathbf{y}^{\alpha},\boldsymbol{\mu}',\mathbb{W}^{\alpha})\neq0
$$
Because the solution $\mathbf{y}$ in the above minimization does not necessarily reproduce the observed equilibrium composition, we should, in addition, evaluate the same quantity at the constrained composition $N^{\alpha}(\mathbf{y}^{\alpha}) = \mathbb{N}^{\alpha}$, leading to the local condition:
\begin{align*}
\min_{\mathbf{y}^{\alpha},C_k^{\alpha},\mathbb{N}^{\alpha}} \Phi_M^{\alpha}(\mathbf{y}^{\alpha},\boldsymbol{\mu}',\mathbb{W}^{\alpha}) &= 
\min_{\mathbf{y}^{\alpha},C_k^{\alpha},\mathbb{N}^{\alpha}} G_M^{\alpha}(\mathbf{y}^{\alpha}, \mathbb{W}^{\alpha}) - \sum_A \mu'_A \mathbb{N}_{A}^{\alpha}
\\ &= \mathbf{G}_M^{\alpha}(\mathbb{N}^{\alpha}, \mathbb{W}^{\alpha}) - \sum_A \mu'_A \mathbb{N}_{A}^{\alpha}
\end{align*}
where the first term is the composition-constrained Gibbs energy. The above two equations define the loss function associated with the observed stable phase. For an unobserved phase $\gamma$, we still require that:
$$
\min_{\mathbf{y}^{\gamma},C_k^{\gamma}} \Phi_M^{\gamma}(\mathbf{y}^{\gamma},\boldsymbol{\mu}',\mathbb{W}^{\gamma}) > 0
$$
with the auxiliary chemical potential $\boldsymbol{\mu}'$. Finally, the total deviation can therefore be measured as follows:
\begin{align*}
\mathcal{L}(\boldsymbol{\mu}',\mathbb{W}) &= \sum_{\alpha\in\text{stable}} \left|\mathbf{G}_m^{\alpha}(\mathbb{X}^{\alpha}, \mathbb{W}^{\alpha}) - \sum_A \mu'_A \mathbb{X}_{A}^{\alpha}\right|^p \\ &+ 
\sum_{\alpha\in\text{stable}} \left|\min_{\mathbf{y}^{\alpha},C_k^{\alpha}} \Phi_m^{\alpha}(\mathbf{y}^{\alpha},\boldsymbol{\mu}',\mathbb{W}^{\alpha})\right|^p \\ &+ \sum_{\gamma\notin\text{stable}} \mathrm{ReLU}\left[-\min_{\mathbf{y}^{\gamma},C_k^{\gamma}} \Phi_m^{\gamma}(\mathbf{y}^{\gamma},\boldsymbol{\mu}',\mathbb{W}^{\gamma})\right]
\end{align*}
at a given auxiliary chemical potential and thermodynamic parameters $\mathbb{W}$. We have normalized the energies to per mole of atoms, and $\mathbb{X}$ is the atomic fraction of the observed phase. The exponent $p=1$ leads to absolute loss, whereas $p=2$ leads to squared loss. In practice, we find that $p=1$ leads to better results because it maintains the magnitude of the loss gradient even if the loss is small. The auxiliary chemical potential in the above equation is unknown and needs to be optimized. Thus, optimal parameters $\mathbb{W}_{\mathrm{opt}}$ can be found as:
$$
\mathbb{W}_{\mathrm{opt}}= \arg\min_{\mathbb{W}} \left[ \sum_{t}\min_{\boldsymbol{\mu}_{t}'} \mathcal{L}_t(\boldsymbol{\mu}_t',\mathbb{W}) + \lambda \|\mathbb{W}\|_2^2 \right]
$$
where we have introduced regularization with weight $\lambda$. The first term includes all target equilibria indexed by $t$ and should be zero when all experimentally observed phase equilibria are reproduced. As the loss function is derived from the equilibrium condition, a value of zero is a necessary and sufficient condition for reproducing the observed phase equilibrium. In practice, $\mathbb{W}_{\mathrm{opt}}$ can be found by minimizing $\mathcal{L}(\boldsymbol{\mu}',\mathbb{W})+ \lambda \|\mathbb{W}\|_2^2$ with respect to $\mathbb{W}$ and $\boldsymbol{\mu}'$ at the same time.

Additional minimizations with respect to internal coordinates appear in the defined loss function for both the grand potential term $\Phi$ and the Gibbs energy $\mathbf{G}$. The internal coordinates $\mathbf{y}^*$ that minimize these quantities are themselves functions of $\mathbb{W}$. Fortunately, using the envelope theorem, it is not necessary to evaluate the derivative of $\mathbf{y}^*$ with respect to $\mathbb{W}$. To calculate the derivative of the loss $\mathcal{L}$ with respect to parameters $\mathbb{W}$, it is only necessary to consider $\mathbf{y}^*$ as input without needing to track the computational graphs that produce $\mathbf{y}^*$. More details regarding the envelope theorem are provided in subsection on envelope theorem. 

\subsection*{Auxiliary Chemical Potential}
\label{auxiliary_potential}

When the number of phases in equilibrium is equal to the number of chemical components, the loss $\sum_{\alpha\in\text{stable}} \left[\mathbf{G}_m^{\alpha}(\mathbb{X}^{\alpha}, \mathbb{W}^{\alpha}) - \sum_A \mu'_A \mathbb{X}_{A}^{\alpha}\right]^2 = 0$ leads to a full-rank linear system from which the auxiliary chemical potential can be solved. For example, in a binary system, knowing the composition and composition-constrained Gibbs energy of two phases allows us to define a chemical-potential tangent plane that passes through both points in the composition-energy space (as shown in Fig.~\ref{fig:loss} a). In such a case, it is possible to define $\boldsymbol{\mu}'$ from the above linear equation so that it is a function of model parameter $\mathbb{W}$, thus eliminating it from optimization together with the corresponding loss terms.

However, this is not possible in general. Consider a two-phase equilibrium in a ternary system. Two points in the energy-composition space cannot uniquely define a chemical-potential tangent plane. However, it is possible to force the tangent plane to cross known points from the two phases in equilibrium. Thus, only one degree of freedom needs to be introduced to define the auxiliary potential, and the first term in the loss function is again eliminated. This is illustrated in Fig.~\ref{fig:auxiliary} and allows us to greatly reduce the number of auxiliary potential terms that need to be minimized in addition to $\mathbb{W}$. The same method can be applied to any number of phases in equilibrium in any number of components. Furthermore, by defining the auxiliary chemical potential, this loss function can also be used without modification when the composition of any phase in equilibrium (e.g. phase $\beta$) is not known. In such cases, we simply require that $\min_{\mathbf{y}^{\beta},C_k^{\beta}} \Phi_m^{\beta}(\mathbf{y}^{\beta},\boldsymbol{\mu}',\mathbb{W}^{\beta}) \to 0$ with respect to $\boldsymbol{\mu}'$ and that $\boldsymbol{\mu}'$ be optimized.

\subsection*{Envelope Theorem}
\label{app_envelop}

The envelope theorem is a well-known result that gives the derivatives of a value function $V$ that has the following form: 
$$
V(\boldsymbol{\omega})=\min_{\mathbf{x}}f(\mathbf{x},\boldsymbol{\omega})
$$
where $\boldsymbol{\omega}$ are parameters. $\mathbf{x}^*(\boldsymbol{\omega})=\arg\min_{\mathbf{x}} f$ is given by the solution to the equation $\partial f(\mathbf{x},\boldsymbol{\omega})/\partial x = 0$, and thus $V(\boldsymbol{\omega}) = f(\mathbf{x}^*(\boldsymbol{\omega}), \boldsymbol{\omega})$. The derivative of $V$ with respect to parameter $\omega$ is then:
$$
\frac{\partial V(\boldsymbol{\omega})}{\partial \omega} = \frac{\partial f}{\partial \omega} + \sum_i \frac{\partial f}{\partial x_i^*}\frac{\partial x_i^*}{\partial \omega} = \frac{\partial f}{\partial \omega}
$$
since $\partial f / \partial x_i^* = 0$ when evaluated at the optimal $\mathbf{x}^*(\boldsymbol{\omega})$. If we have constraints on $\mathbf{x}$ given by $g(\mathbf{x},\boldsymbol{\omega})=0$, $\mathbf{x}^*(\boldsymbol{\omega})$ is determined by the stationary point of the Lagrangian:
$$
\frac{\partial f}{\partial x} + \lambda \frac{\partial g}{\partial x} = 0; \quad g=0
$$
Since the constraints are always satisfied at $\mathbf{x}^*(\boldsymbol{\omega})$,
$$
\frac{\partial g(\mathbf{x}^*(\boldsymbol{\omega}),\boldsymbol{\omega})}{\partial \omega} = \frac{\partial g}{\partial \omega} + \sum_i \frac{\partial g}{\partial x_i^*}\frac{\partial x_i^*}{\partial \omega} \equiv 0 
$$
The derivative is:
\begin{align*}
\frac{\partial V(\boldsymbol{\omega})}{\partial \omega} &= \frac{\partial f}{\partial \omega} + \sum_i \frac{\partial f}{\partial x_i^*}\frac{\partial x_i^*}{\partial \omega} \\
&= \frac{\partial f}{\partial \omega} + \sum_i \frac{\partial f}{\partial x_i^*}\frac{\partial x_i^*}{\partial \omega} + \lambda \left[ \frac{\partial g}{\partial \omega} + \sum_i \frac{\partial g}{\partial x_i^*}\frac{\partial x_i^*}{\partial \omega} \right]
\end{align*}
which is equal to $\partial L/\partial \omega$ if $g$ explicitly depends on $\omega$ and $\partial f/\partial \omega$ otherwise, both evaluated at $\mathbf{x}^*(\boldsymbol{\omega})$. In our loss function, terms have the form $\min_{\mathbf{y}} G(\mathbf{y},\mathbb{W})$ and the envelope theorem can be readily applied to these terms to calculate their derivative with respect to $\mathbb{W}$. Finally, the derivative of the loss function with respect to $\mathbb{W}$ can be calculated using the chain rule. Thus, full differentiation through the determination of $\mathbf{y}^*$ can be avoided, reducing memory usage and computational cost.

\subsection*{Thermodynamic Model}
\label{calphad_models}

To describe a thermodynamic system, it is necessary for the model to provide free energies for different possible phases. In the CALPHAD approach, each phase has its own thermodynamic model with different parameters, and the thermodynamic system is an ensemble of defined models. However, it is also possible for a global model to provide the thermodynamic description of all phases, as may be the case for machine learning models. 

We use CALPHAD models as examples of the optimization. In CALPHAD, the free energy of the phase is commonly described by the compound energy formalism (CEF), in which all sites in the unit cell are partitioned into sublattices. We define the site fraction for sublattice $s$ of element $A$ as $y_A^{(s)} = n_A^{(s)}/(n_A^{(s)}+n_B^{(s)}+\cdots)$
where $n_A^{(s)}$ is the number of atoms of element $A$ per molar formula on sublattice $s$. \cite{lukasComputationalThermodynamicsCalphad2007} The total number of atoms of component $A$ in a molar formula unit is obtained by counting the number of $A$ on each sublattice, $N_i = \sum_s N_s y_i^{(s)}$, where $N_s$ is the number of sites in sublattice $s$ in a molar formula unit. Omitting pressure, the thermodynamic model is given by:
$$
G_M(\mathbf{y},T) = \sum_{I} P_{I}(\mathbf{y}) g_I + RT \sum_s N_s \sum_{i=A}^{N} y_i^{(s)}\ln y_i^{(s)} + G_M^{\mathrm{ex}}(\mathbf{y},T)
$$
where the first sum over $I$ is over possible component arrays specifying the occupancy of sites in the end-members. For example, $I=(AB\cdots)$ with $A$ occupying the first sublattice, etc., and $g_I$ can be interpreted as the free energy of the end-member configuration $g_{AB\cdots}$, which can be written as temperature polynomials. The value of the coefficient $P_I$ is $y_A^{(1)}y_B^{(1)}\cdots$.

The excess energy can consist of different terms. Contributions from pairwise and ternary excess terms on the same sublattice, as well as binary mixing on two sublattices, are given by Lukas et al. \cite{lukasComputationalThermodynamicsCalphad2007}, where $L^{(n)}=L(T)$ are temperature polynomials indexed by integer $n$:
\begin{widetext}
\begin{align*}
G^{\mathrm{ex,pair}}_{M,ab\cdots\underbrace{(ij)}_{(s)}\cdots c}
=& y_a^{(1)}y_b^{(2)}\cdots (y_i^{(s)}y_j^{(s)})\cdots y_c^{(N)} \left[
\sum_{n=0}^{v}L^{(n)} (y_i^{(s)}-y_j^{(s)})^n\right] \\
G^{\mathrm{ex,ternary}}_{M,ab\cdots\underbrace{(ijk)}_{(s)}\cdots c}
=& y_a^{(1)}y_b^{(2)}\cdots (y_i^{(s)}y_j^{(s)}y_k^{(s)})\cdots y_c^{(N)} \left[
v_i L^{(0)} + v_j L^{(1)} + v_k L^{(2)} \right]  \\
G^{\mathrm{ex,2pair}}_{M,ab\cdots\underbrace{(ij)}_{(s)}\cdots\underbrace{(mn)}_{(r)}\cdots c}
=& y_a^{(1)}y_b^{(2)}\cdots (y_i^{(s)}y_j^{(s)})\cdots(y_m^{(r)}y_n^{(r)})\cdots y_c^{(N)} \times \\ & \quad \left[
L^{(0)} + (y_m^{(r)}-y_n^{(r)})L^{(1)} + (y_i^{(s)}-y_j^{(s)})L^{(2)} 
\right]
\end{align*}
\end{widetext}
where the index $ab\cdots(ij)_{(s)}\cdots c$ means that the specific term is related to mixing on the $(s)$-th sublattice with components $i$ and $j$, while all other sublattices are occupied by $a,b,\cdots, c$, respectively. The same notation can be extended to ternary mixing and mixing on different sublattices. In ternary mixing, the normalized term $v_i = y^{(s)}_i + (1-\sum_j y^{(s)}_j)/3$ is used. The compound energy formalism covers solid solutions (single sublattice), intermetallics (multiple sublattices) and compounds (multiple sublattices with fixed site fractions).

To calculate the loss function, the minimized Gibbs energy or grand potential must be calculated. In the compound energy formalism, the minimization problem has the following constraints on $\mathbf{y}$: $0 \le \mathbf{y} \le 1$ and $\sum_i y_i^{(s)} = 1$. For Gibbs energy, composition constraints $\mathbb{X}_i = N_i/ \sum_i N_i$ also have to be imposed. Explicitly, we have $(N-1)$ composition constraints for $N$ chemical species:
\begin{align*}
\sum_s N_s y_{i}^{(s)} &= \mathbb{X}_A \left(\sum_{j\neq \mathrm{Va}}\sum_t N^{(t)} y_{j}^{(t)}\right) \\
&= \mathbb{X}_A \left(\sum_s N^{(s)}-\sum_t N^{(t)} y_{\mathrm{Va}}^{(t)} \right) 
\end{align*}
which is linear. Our approach to calculating terms in the loss function consists of three steps. First, a random grid of feasible points $\mathbf{y}_i$ is prepared. This can be done using Dirichlet sampling when calculating the grand potential. When sampled points need to satisfy imposed composition constraints, the feasible points lie in a polytope given by the above constraint equations, and hit-and-run sampling can be used.

Second, for each sampled point, gradient descent optimization is performed for $N$ steps. In particular, the exponential gradient descent (EGD) method can conveniently account for the constraints on the internal coordinates $\mathbf{y}$ (Detailed in the subsection on exponential gradient descent). Denoting the linear constraints as $\mathbf{A} \mathbf{y} - \mathbf{B} = 0$, the $t+1$ step of $\mathbf{y}^{(t+1)}$ in EGD can be determined by finding the stationary point of the Lagrangian:
\begin{align*}
L = & \eta \sum_{is} g_{is}^{(t)} y_{is}^{(t+1)} + \sum_{is} y_{is}^{(t+1)} \log\left(\frac{y_{is}^{(t+1)}}{y_{is}^{(t)}}\right) \\
&+ \sum_s \lambda_s \left(\sum_{i} y_{is}^{(t+1)} - 1\right) + \boldsymbol{\mu} (\mathbf{A} \mathbf{y}^{(t+1)} - \mathbf{B})
\end{align*}
where $\mathbf{g}^{(t)}$ and $\mathbf{y}^{(t)}$ are the gradient and coordinate vector at step $t$, and $\boldsymbol{\mu}$ is a vector of $N-1$ multipliers. $\lambda_s$ is the multiplier for each sublattice. The stationary point is given by:
$$
y_{is}^{(t+1)} = \frac{y_{is}^{(t)} e^{-\eta g_{is}^{(t)} - \sum_a\mu_a A_{a,is}}}{e^{1+\lambda_s}}
$$
The term $e^{1+\lambda_s}$ is the normalizer for each sublattice, and thus $\lambda_s$ can be eliminated to give:
$$
y_{is}^{(t+1)} = \frac{y_{is}^{(t)} e^{-\eta g_{is}^{(t)} - \sum_a\mu_a A_{a,{is}}}}{\sum_{j} y_{js}^{(t)} e^{-\eta g_{js}^{(t)} - \sum_a\mu_a A_{a,js}}}
$$
Without composition constraints, the update is simply given by $y_{is}^{(t+1)} = y_{is}^{(t)} e^{-\eta g_{is}^{(t)}}/\sum_{j} y_{js}^{(t)} e^{-\eta g_{js}^{(t)}}$. With constraints, the multipliers $\boldsymbol{\mu}$ are determined by the now nonlinear set of composition-constraint equations $\mathbf{A} \mathbf{y}^{(t+1)} - \mathbf{B} = 0$ and need to be solved numerically. This is a much smaller problem, and the numerical solution can be efficiently found by Newton's method with an analytic Jacobian matrix. When the multipliers are solved, $y_{is}^{(t+1)}$ can be calculated accordingly. Such constrained optimization can be efficiently done in parallel for all sampled points. By choosing a relatively dense sample grid, a small number of gradient descent steps is sufficient to ensure that a global minimum can be found. We note that the formula of EGD is similar to Kikuchi's natural iteration method (NI) in the cluster variation method (CVM), \cite{kikuchiTernaryPhaseDiagram1977} which is fundamentally due to the interpretation of internal variables as probability distributions in both cases.

Finally, all $\mathbf{y}'$ obtained after gradient descent are used as inputs to compute the Gibbs energy and grand potential. All sampled points are considered in calculating the minimum by using a $\mathrm{Softmin}$ function:
$$
\min_{\mathbf{y}} F (\mathbf{y}, \mathbb{W}) = -\tau \log \sum_j \exp \left[-\frac{F (\mathbf{y}_j, \mathbb{W})}{\tau}\right]
$$
where $F$ is either $G_m(\mathbf{y}, \mathbb{W})$ or $\Phi_m(\mathbf{y}, \boldsymbol{\mu}, \mathbb{W})$.
One benefit of using this formula is that all coordinates that yield degenerate energies are included, thus avoiding discontinuity from choosing any one of them. The results of $\mathrm{Softmin}$ are then inserted into the loss function. While $\mathrm{Softmin}$ is an approximation, its error can be controlled by the parameter $\tau$ to within $1\,\mathrm{J/mol}$ by setting $\tau=1/\log(n)$, where $n$ is the total number of samples.

\subsection*{Exponential Gradient Descent with Constraints}
\label{app_egd}

Ordinary gradient descent, $\mathbf{x}^{t+1} = \mathbf{x}^{t} - \eta \mathbf{g}^{(t)}$, where $\mathbf{g}^{(t)}$ is the gradient of the target function $f$ at step $t$, can be equivalently expressed as the solution to a minimization problem:
\begin{align*}
\mathbf{x}^{t+1} &= \arg\min_{\mathbf{x}} \left\{ \underbrace{f(\mathbf{x}^{(t)}) + \langle \mathbf{g}, \mathbf{x}-\mathbf{x}^{(t)}\rangle}_{\text{linearized estimate at $\mathbf{x}$}} + \underbrace{\frac{\lambda}{2}||\mathbf{x}-\mathbf{x}^{(t)}||^2}_{\text{penalty}}\right\} \\
&= \arg\min_{\mathbf{x}} \left\{ \eta \langle \mathbf{g}, \mathbf{x}\rangle + \frac{1}{2}||\mathbf{x}-\mathbf{x}^{(t)}||^2\right\}
\end{align*}
The penalty term ensures that the linear approximation remains valid. The second line is obtained by removing the part that does not depend on $\mathbf{x}$ and scaling by $\eta=1/\lambda$. $\langle a,b\rangle$ is the inner product between vectors $a$ and $b$. Denoting the bracketed term by $L$, minimizing $L$ leads to:
$$
\frac{\partial L}{\partial x_i^{(t+1)}} = \eta g_i^{(t)} + x_i^{(t+1)} - x_i^{(t)}= 0
$$

Similarly, we can change the penalty term from the Euclidean norm to a divergence measure to reflect the property of the underlying geometry. In the simplex, a vector can be interpreted as a probability distribution, and based on this, the Kullback-Leibler (KL) divergence measure can be introduced:
$$
\mathbf{x}^{t+1} = \arg\min_{\mathbf{x}\in\mathcal{X}} \left\{ \eta \langle \mathbf{g}^{(t)}, \mathbf{x}\rangle + D_{\mathrm{KL}}(\mathbf{x}|\mathbf{x}^{(t)})\right\}
$$
and we have to constrain $\mathbf{x}$ to be in the simplex: $\mathcal{X} = \{\mathbf{x}\mid \mathbf{x} \ge 0, \sum_i x_i = 1\}$. The constrained minimization can be performed using the Lagrangian:
$$
L = \eta \sum_i g_i^{(t)} x_i + \sum_i x_i^{(t+1)} \log\left(\frac{x_i}{x_i^{(t)}}\right) + \lambda \left(\sum_ix_i^{(t+1)} - 1\right)
$$
and the solution is given by:
\begin{gather*}
x_i^{(t+1)} = x_i^{(t)} e^{-\eta g_i^{(t)} - 1 - \lambda} \\
e^{-1-\lambda} \sum_i x_i^{(t)} e^{-\eta g_i^{(t)}} = 1  
\end{gather*}
so that we see $e^{-1-\lambda}$ is just a normalization factor, leading to:
$$
x_i^{(t+1)} = \frac{x_i^{(t)} e^{-\eta g_i^{(t)}}}{\sum_i x_i^{(t)} e^{-\eta g_i^{(t)}}}
$$

As described previously, the above can be extended to multiple sublattices and linear constraints, as is the case in CEF. In the following, we describe the solution to the $N-1$ multiplier $\boldsymbol{\mu}$ that appears due to the composition constraints $\mathbf{A} \mathbf{y}^{(t+1)} - \mathbf{B}=0$. For a constraint indexed by $b$:
$$
F_b(\boldsymbol{\mu}) = \sum_{is} A_{b,is} y_{is}^{(t+1)} - B_b = 0
$$
Newton's method can be used to iteratively solve the non-linear equations. Using $\boldsymbol{\mu}'$ to denote current value of $\boldsymbol{\mu}$, we find:
$$
F_b(\boldsymbol{\mu}') + \sum_{a} \frac{\partial F_b(\boldsymbol{\mu}')}{\partial \mu_{a}} \Delta \mu_{a} = 0
$$
leading to
$$
\mathbf{F}(\boldsymbol{\mu}') + \mathbf{J} (\boldsymbol{\mu}') \Delta \boldsymbol{\mu} = 0
$$
where $\mathbf{J}$ is the Jacobian with matrix elements $\mathbf{J}_{ab} = \partial F_a / \partial \mu_b$. The Newton update can be found by $\Delta \boldsymbol{\mu} = -\mathbf{J}^{-1} (\boldsymbol{\mu}') \mathbf{F}(\boldsymbol{\mu}')$.
The Jacobian matrix elements are given by:
$$
\frac{\partial F_b}{\partial \mu_a} = \sum_{is} A_{b,is} \frac{\partial y_{is}^{(t+1)}}{\partial \mu_a} \\
\frac{\partial y_{is}^{(t+1)}}{\partial \mu_a} = y_{is}^{(t+1)} \left( \sum_j A_{a,js} y^{(t+1)}_{js} - A_{a,is}\right)
$$
where it should be noted that $\mathbf{y}^{(t+1)}$ is a function of the current $\boldsymbol{\mu}'$.

\section*{Conflict of Interest}
The authors have no conflicts to disclose.

\section*{Author Contributions}
\textbf{Wenhao Zhang}: Conceptualization, Data Curation, Formal Analysis, Investigation, Methodology, Writing of the original draft, review and editing.
\textbf{Yusuke Matsuoka}: Methodology, review and editing.
\textbf{Jean-Claude Crivello}: Writing (review and editing).
\textbf{Koyama Toshiyuki}: Supervision, Writing (review and editing).
\textbf{Taichi Abe}: Supervision, Writing (review and editing).

\section*{Data Availability Statement}

The data that support the findings of this study, including the computational notebook and results, are available at \url{https://github.com/whzhangg/eqoptimizer}.


\bibliography{phopt}

\clearpage
\onecolumngrid

\section*{Supplemental Material}

\setcounter{figure}{0}
\renewcommand{\thefigure}{S\arabic{figure}}

\subsection*{Target phase equilibria for simplified Au--Cu}

\begin{figure*}[htb]
\centering
\includegraphics[width=0.7\linewidth]{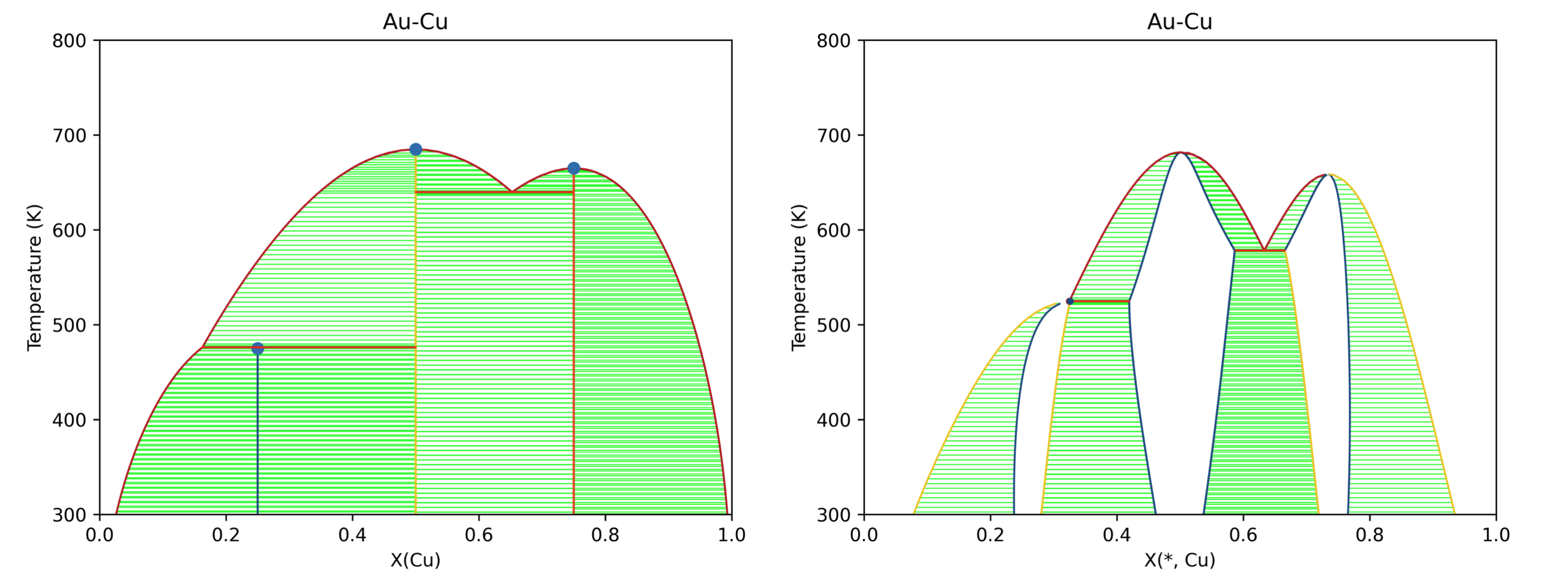}
\caption{Left: simplified order-disorder phase diagram used to generate the target phase equilibria for optimization in Au--Cu binary. The three points indicate the transition temperature of the three ordered phase: $475$, $685$ and $665\,\mathrm{K}$. Right: the CALPHAD assessment by Sundman using a four-sublattice split-CEF model.}
\label{fig:aucu_target}
\end{figure*}

In the simplified Au--Cu binary phase diagram, we have considered the disordered FCC ideal solid solution phase and three ordered phase approximated as line compounds. In this simplified case, the phase competition between ordered and disordered phases are attributed to only the enthalpy difference between the ordered phase and solid solution: $\Delta H_{\mathrm{AuCu_3}}$, $\Delta H_{\mathrm{Au_3Cu}}$ and $\Delta H_{\mathrm{AuCu}}$ while transitions from ordered phases to the disordered phase occur at temperature:
$$
T = \frac{\Delta H}{R(x_{\mathrm{Au}} \ln x_{\mathrm{Au}} + x_{\mathrm{Cu}} \ln x_{\mathrm{Cu}})}
$$
Using the transition temperature, the enthalpy difference $\Delta H_{\mathrm{AuCu_3}}$ and $\Delta H_{\mathrm{AuCu}}$ are determined to be $-3109$ ($665\,\mathrm{K}$) and $-3947\,\mathrm{J/mol}$ ($685\,\mathrm{K}$), respectively. For the transition related to the $\mathrm{Au_3Cu}$ phase, $\Delta H_{\mathrm{Au_3Cu}}$ is set to $-2325\,\mathrm{J/mol}$ so that the ordered phase is stable up to $475\,\mathrm{K}$. The solid-state phase diagram from these values are plotted in Fig.~\ref{fig:aucu_target}. It should be noted that the phase diagram does not represent experimental results, and is used for the illustration of the optimization.

\subsection*{Thermodynamic models to be optimized}

\subsubsection*{Al--Zn}

In the binary system of Al--Zn, three phases are modelled: the liquid phase, the FCC phase and the HCP phase. For the liquid and the FCC phase, the Gibbs energy CEF model considers a single sublattice with mixing of both elements:
$$
G(\mathbf{x},T) = x_{\mathrm{Al}} G_{\mathrm{Al}} + x_{\mathrm{Zn}} G_{\mathrm{Zn}} + RT \sum_{i=\mathrm{Al},\mathrm{Zn}} x_i \ln x_i + x_{\mathrm{Al}}x_{\mathrm{Zn}} \left[L_0 + L_1 (x_{\mathrm{Al}}-x_{\mathrm{Zn}})\right]
$$
where $G_{\mathrm{Al}}$, $G_{\mathrm{Zn}}$ are end-member energy and $L_0$, $L_1$ are interaction parameters. Both $G$ and $L$ are polynomials of temperature. As described in the main text, optimization is with respect to the first order correction $a+bT$ to these parameters. For the HCP phase, considering its limited homogeneity range, its thermodynamic model omits the $L_1$ term:
$$
G(\mathbf{x},T) = x_{\mathrm{Al}} G_{\mathrm{Al}} + x_{\mathrm{Zn}} G_{\mathrm{Zn}} + RT \sum_{i=\mathrm{Al},\mathrm{Zn}} x_i \ln x_i + x_{\mathrm{Al}}x_{\mathrm{Zn}} L_0 
$$
In total, optimization contains 22 parameters. 

\subsubsection*{Cu--Mg}

In the binary system Cu--Mg, nine phases are modelled. The liquid phase are modelled including the $L_0$ and $L_1$ parameters, which is same as the liquid and FCC phase in Al--Zn binary. The two solid solution phase FCC and HCP considers the $L_0$ parameter. Mg$_2$Cu,  Mg$_5$Cu, MgCu, MgCu$_3$, MgCu$_4$ are modelled as line compounds (single end-member) with two sublattices. Their free energies $G_{\mathrm{Mg_2Cu}}$, $G_{\mathrm{Mg_5Cu}}$, $G_{\mathrm{MgCu}}$, $G_{\mathrm{MgCu_3}}$ and $G_{\mathrm{MgCu_4}}$ are again given by temperature dependent polynomials to be used in the thermodynamic modeling. For the Laves phase, its two sublattice CEF model is given as a function of site-fractions:
\begin{align*}
G(\mathbf{y},T) = & y_{\mathrm{Mg}}^{(1)}y_{\mathrm{Mg}}^{(2)} G_{\mathrm{MgMg_2}} 
+ y_{\mathrm{Mg}}^{(1)}y_{\mathrm{Cu}}^{(2)} G_{\mathrm{MgCu_2}}
+ y_{\mathrm{Cu}}^{(1)}y_{\mathrm{Mg}}^{(2)} G_{\mathrm{CuMg_2}}
+ y_{\mathrm{Cu}}^{(1)}y_{\mathrm{Cu}}^{(2)} G_{\mathrm{CuCu_2}} \\
& \quad + RT \sum_{s=1,2} N_s \sum_{i=\mathrm{Cu},\mathrm{Mg}} y_i^{(s)}\ln y_i^{(s)}
\end{align*}
where $y_i^{(s)}$ is the site fraction of element $i$ on sublattice $s$. For the Laves phase, two sublattices have multiplicity ratio $1:2$. $G_{\mathrm{AB_2}}$ are the end-member energies. The optimization of the Laves phase is done by considering $a+bT$ terms to the end-member energies.

\subsubsection*{Al--Mo--Si}

In the ternary Al--Mo--Si system, the thermodynamic models are taken from the previous assessment. 14 phases are considered in total. Among them, six are modelled as line compounds: Al$_{12}$Mo, Al$_5$Mo, Al$_4$Mo, Al$_8$Mo$_3$, Al$_{63}$Mo$_{37}$ and Mo$_{0.333}$Al$_{0.233}$Si$_{0.434}$ (denoted as the $C54$) phase. The optimization of these line compounds is performed by $a+bT$ terms to their energy. 

The Si diamond phase is described by a binary solid solution with Si and Al elements, given by:
$$
G(\mathbf{x},T) =  x_{\mathrm{Al}} G_{\mathrm{Al}} + x_{\mathrm{Si}} G_{\mathrm{Si}} + RT \sum_{i=\mathrm{Al},\mathrm{Si}} x_i \ln x_i + x_{\mathrm{Al}} x_{\mathrm{Si}} L_0
$$

There are three ternary solid solution phases: BCC phase, FCC phase and the liquid phase. These phases are modelled by:
$$
G(\mathbf{x},T) =  x_{\mathrm{Al}} G_{\mathrm{Al}} + x_{\mathrm{Mo}} G_{\mathrm{Mo}} + x_{\mathrm{Si}} G_{\mathrm{Si}} + RT \sum_{i=\mathrm{Al},\mathrm{Mo},\mathrm{Si}} x_i \ln x_i + G^{\mathrm{excess}}
$$
where for the BCC phase:
\begin{align*}
G^{\mathrm{excess}} =  & x_{\mathrm{Mo}}x_{\mathrm{Si}} L_0^{(\mathrm{Mo,Si})} \\
& + x_{\mathrm{Al}}x_{\mathrm{Mo}} \left[L_0^{(\mathrm{Al,Mo})} + L_1^{(\mathrm{Al,Mo})} (x_{\mathrm{Al}}-x_{\mathrm{Mo}}) + L_2^{(\mathrm{Al,Mo})} (x_{\mathrm{Al}}-x_{\mathrm{Mo}})^2\right]
\end{align*}
with four $L$ interaction parameters for the excess Gibbs energy. For the FCC phase:
$$
G^{\mathrm{excess}} =  x_{\mathrm{Al}}x_{\mathrm{Mo}} \left[L_0^{(\mathrm{Al,Mo})} + L_1^{(\mathrm{Al,Mo})} (x_{\mathrm{Al}}-x_{\mathrm{Mo}})\right]   + x_{\mathrm{Al}}x_{\mathrm{Si}} L_0^{(\mathrm{Al,Si})}
$$
and for the liquid phase:
\begin{align*}
G^{\mathrm{excess}} &=  x_{\mathrm{Al}}x_{\mathrm{Mo}} \left[L_0^{(\mathrm{Al,Mo})} + L_1^{(\mathrm{Al,Mo})} (x_{\mathrm{Al}}-x_{\mathrm{Mo}}) + L_2^{(\mathrm{Al,Mo})} (x_{\mathrm{Al}}-x_{\mathrm{Mo}})^2\right] \\
& + x_{\mathrm{Al}}x_{\mathrm{Si}} \left[L_0^{(\mathrm{Al,Si})} + L_1^{(\mathrm{Al,Si})} (x_{\mathrm{Al}}-x_{\mathrm{Si}}) + L_2^{(\mathrm{Al,Si})} (x_{\mathrm{Al}}-x_{\mathrm{Si}})^2\right] \\
& + x_{\mathrm{Mo}}x_{\mathrm{Si}} \left[L_0^{(\mathrm{Mo,Si})} + L_1^{(\mathrm{Mo,Si})} (x_{\mathrm{Mo}}-x_{\mathrm{Si}}) + L_2^{(\mathrm{Mo,Si})} (x_{\mathrm{Mo}}-x_{\mathrm{Si}})^2\right]
\end{align*}
where each of the $L$ parameters are optimized with $a+bT$ corrections. There are four intermetallic phases with multiple end-members. Three of them: $\mathrm{Mo_3(AlSi)}$ and two $\mathrm{Mo(AlSi)_2}$ phases denoted as C40 and C11B are pseudo-binary with Mo occupying the first sublattice and Al--Si mixing on the second sublattice. Their free energy is given by:
$$
G(\mathbf{y},T) =  y_{\mathrm{Al}}^{(2)} G_{\mathrm{MoAl}} + y_{\mathrm{Si}}^{(2)} G_{\mathrm{MoSi}} + RT \sum_{i=\mathrm{Al},\mathrm{Si}} y_i^{(2)} \ln y_i^{(2)} + G^{\mathrm{excess}}
$$
where the $G_{AB}$ denote the endmember with the appropriate sublattice multiplicity. The excess term is 
$$
G^{\mathrm{excess}} = y_{\mathrm{Al}}^{(2)} y_{\mathrm{Si}}^{(2)} L_0^{\mathrm{(Mo:Al,Si)}}
$$
for the $\mathrm{Mo_3(AlSi)}$  and the $\mathrm{Mo(AlSi)_2}$ C11B phase, and 
$$
G^{\mathrm{excess}} = y_{\mathrm{Al}}^{(2)} y_{\mathrm{Si}}^{(2)} \left[L_0^{\mathrm{(Mo:Al,Si)}} + L_1^{\mathrm{(Mo:Al,Si)}}(y_{\mathrm{Al}}^{(2)}-y_{\mathrm{Si}}^{(2)})\right]
$$
for the $\mathrm{Mo(AlSi)_2}$ C40 phase.

For the ternary intermetallic phase derived from $\mathrm{Mo_5Si_3}$, the thermodynamic model considers mixing of Mo and Si on the first sublattice and all three elements on the second sublattice. Its CEF expression is given by:
\begin{align*}
G(\mathbf{y},T) = & 
y_{\mathrm{Mo}}^{(1)}y_{\mathrm{Mo}}^{(2)} G_{\mathrm{MoMo}} +  y_{\mathrm{Mo}}^{(1)}y_{\mathrm{Al}}^{(2)} G_{\mathrm{MoAl}} + y_{\mathrm{Mo}}^{(1)}y_{\mathrm{Si}}^{(2)} G_{\mathrm{MoSi}} \\ 
& + y_{\mathrm{Si}}^{(1)}y_{\mathrm{Mo}}^{(2)} G_{\mathrm{SiMo}} +  y_{\mathrm{Si}}^{(1)}y_{\mathrm{Al}}^{(2)} G_{\mathrm{SiAl}} + y_{\mathrm{Si}}^{(1)}y_{\mathrm{Si}}^{(2)} G_{\mathrm{SiSi}} \\
& + y_{\mathrm{Mo}}^{(1)}y_{\mathrm{Si}}^{(1)}y_{\mathrm{Mo}}^{(2)} L_0^{\mathrm{Mo,Si:Mo}} 
+ y_{\mathrm{Mo}}^{(1)}y_{\mathrm{Si}}^{(1)}y_{\mathrm{Si}}^{(2)} L_0^{\mathrm{Mo,Si:Si}} \\
& + y_{\mathrm{Si}}^{(1)}y_{\mathrm{Mo}}^{(2)}y_{\mathrm{Si}}^{(2)} L_0^{\mathrm{Si:Mo,Si}} 
+ y_{\mathrm{Mo}}^{(1)}y_{\mathrm{Mo}}^{(2)}y_{\mathrm{Si}}^{(2)} L_0^{\mathrm{Mo:Mo,Si}} \\
& + y_{\mathrm{Mo}}^{(1)}y_{\mathrm{Al}}^{(2)}y_{\mathrm{Si}}^{(2)} L_0^{\mathrm{Mo,Al:Si}} \\
& + RT \left[ N_1 \sum_{i=\mathrm{Mo,Si}} y_i^{(1)}\ln y_i^{(1)} + N_2 \sum_{i=\mathrm{Mo,Al,Si}} y_i^{(2)}\ln y_i^{(2)}\right]
\end{align*}
with a total of six end-member energies and five interaction parameters to be optimized.

\end{document}